\documentclass[preprint,tightenlines,showpacs,amsmath,amssymb]{revtex4}
\usepackage{graphicx}% Include figure files
\usepackage{dcolumn}% Align table columns on decimal point
\usepackage{bm}% bold math
\usepackage{psfig}
\usepackage{rotating}
\begin{document}
\normalsize
\parskip=5pt plus 1pt minus 1pt

\title{\boldmath Measurements of $J/\psi$ decays into $\omega K \bar{K} \pi$, $\phi K \bar{K} \pi$ and $\eta K^{0}_{S} K^{\pm} \pi^{\mp}$}
\author{
\vspace{0.2cm}
%(China and Univ. Hawaii)\\
\vspace{0.3cm}
M.~Ablikim$^{1}$,              J.~Z.~Bai$^{1}$,   Y.~Bai$^{1}$,
Y.~Ban$^{11}$,
X.~Cai$^{1}$,                  H.~F.~Chen$^{16}$,
H.~S.~Chen$^{1}$,              H.~X.~Chen$^{1}$, J.~C.~Chen$^{1}$,
Jin~Chen$^{1}$
% ...
\\
\vspace{0.2cm}
(BES Collaboration)\\
\vspace{0.2cm}
{\it
$^{1}$ Institute of High Energy Physics, Beijing 100049, People's Republic of China\\
% ...
$^{11}$ Peking University, Beijing 100871, People's Republic of China\\
% ...
$^{16}$ University of Science and Technology of China, Hefei 230026,
People's Republic of China\\
% ...
}}
%\end{center}
\vspace{0.4cm}
\date{\today}

\begin{abstract}

The decays of $J/\psi \rightarrow \omega K\bar{K}\pi$ and $J/\psi
\rightarrow \phi K\bar{K}\pi$ are studied using $5.8 \times 10^{7}$
$J/\psi$ events collected with the Beijing Spectrometer (BESII) at the
Beijing Electron-Positron Collider (BEPC). The
$K^{0}_{S}K^{\pm}\pi^{\mp}$ and $K^{+}K^{-}\pi^{0}$ systems, produced
in $J/\psi \rightarrow \omega K\bar{K}\pi$, have enhancements in the
invariant mass distributions at around $1.44$ GeV/$c^{2}$.  However,
there is no evidence for mass enhancements in the $K\bar{K}\pi$ system
in $J/\psi \rightarrow \phi K\bar{K}\pi$.  The branching fractions of
$J/\psi \rightarrow \omega K^{0}_{S}K^{\pm}\pi^{\mp}$, $\phi
K^{0}_{S}K^{\pm}\pi^{\mp}$, $\omega K^{*}\bar{K}+c.c.$, and $\phi
K^{*}\bar{K}+c.c.$ are obtained, and the
$J/\psi \rightarrow \eta K^{0}_{S}K^{\pm}\pi^{\mp}$ branching fraction
is measured for the first time.

\end{abstract}


\maketitle

\section{\bf INTRODUCTION}

A pseudoscalar gluonium candidate, the so-called $E/\iota(1440)$, was
observed in $p\bar{p}$ annihilation in 1967~\cite{baillon67} and in
$J/\psi$ radiative decays in the
1980's~\cite{scharre80,edwards82e,augustin90}. After 1990, more and
more observations revealed the existence of two resonant structures
around 1.44 GeV/$c^{2}$ in $a_0(980)\pi$, $K\bar{K}\pi$, and
$K^*\bar{K}$
spectra~\cite{rath89,adams01,bai90c,augustin92,bertin95-97,cicalo99,nichitiu02}.
They showed that the lower state, $\eta(1405)$, has large couplings to
$a_0(980)\pi$ and $K\bar{K}\pi$, while the high mass state,
$\eta(1475)$, favors $K^*\bar{K}$. The $\eta(1405)$ was also confirmed
by MarkIII~\cite{bolton92b}, Crystal Barrel~\cite{amsler95f}, and
DM2~\cite{augustin90} in decays into $\eta\pi\pi$ in $J/\psi$
radiative decays and $\bar{p}p$ annihilations.

In contrast, although $\eta(1475)$ was observed in $K\bar{K}\pi$
($K^*\bar{K}$)~\cite{rath89,adams01,bai90c,augustin92,bertin95-97,cicalo99,nichitiu02},
it was not seen in $\eta\pi\pi$. Nonetheless, the study of
$K\bar{K}\pi$ and $\eta\pi\pi$ channels in $\gamma\gamma$
collisions~\cite{acciarri01g} showed that $\eta(1475)$ appeared in
$K\bar{K}\pi$, but not in $\eta\pi\pi$, while $\eta(1405)$ appeared in
neither channel.

The study of the decays $J/\psi \rightarrow
\{\gamma,\omega,\phi\}K\bar{K}\pi$ is a useful tool in the
investigation of quark and possible gluonium content of the states around
1.44 GeV/$c^{2}$.  In this paper, we investigate the possible
structure in the $K\bar{K}\pi$ final state in $J/\psi$ hadronic decays
at around $1.44$ GeV/$c^{2}$, and measure the branching fraction of $J/\psi \rightarrow \eta K^{0}_{S} K^{\pm} \pi^{\mp}$ for the first time, based on $5.8 \times 10^{7}$ $J/\psi$
events collected with the Beijing Spectrometer at the Beijing
Electron-Positron Collider (BEPC) .

\section{\bf THE BES DETECTOR}

BESII is a large solid-angle magnetic
spectrometer that is described in detail elsewhere~\cite{jzbnpa}.
 Charged particle momenta
are determined with a resolution of $\sigma_{p}/p=1.78\%\sqrt{1+p^{2}}$ ($p$ in GeV/$c^{2}$) in a
40-layer cylindrical main drift chamber (MDC). Particle identification (PID) is
accomplished using specific ionization ($dE/dx$) measurement in the
MDC and time-of-flight (TOF) information in a barrel-like
array of 48 scintillation counters. The $dE/dx$ resolution is
$\sigma_{dE/dx} \simeq 8.0\%$; the TOF resolution is
$\sigma_{TOF}=180$ ps for Bhabha events. Outside of the
time-of-flight counter is a 12-radiation-length barrel shower
counter (BSC) comprised of gas proportional tubes interleaved with
lead sheets. The BSC measures the energy and direction of photons
with resolutions of $\sigma_{E}/E \simeq 21\%\sqrt{E}$ ($E$ in GeV),
$\sigma_{\phi}=7.9$ mrad, and $\sigma_{z}=2.3$ cm. The iron flux
return of the magnet is instrumented with three double layers of
counters that are used to identify mouns.

 A Geant3 based Monte Carlo (MC) package (SIMBES) with detailed consideration of the detector performance is used.
 The consistency between data and MC has been carefully checked in many
high purity physics channels, and the agreement
 is reasonable~\cite{bessimulation2005}. The detection efficiencies and mass resolutions for
 each decay mode presented in this paper are obtained with uniform phase space MC generators.

\section{\bf ANALYSIS}

In this analysis, $\omega$ mesons are observed in the $\omega \rightarrow \pi^{+}\pi^{-}\pi^{0}$ decay, $\phi$ mesons in the $\phi \rightarrow K^{+}K^{-}$
decay, and other mesons are detected in the decays: $K^{0}_{S}
\rightarrow \pi^{+}\pi^{-}$, $\pi^0 \rightarrow \gamma \gamma$, $\eta
\rightarrow \pi^{+}\pi^{-}\pi^{0}$. The final states of the analyzed
decays $J/\psi \rightarrow \{\omega,\eta\} K^{0}_{S}K^{\pm}
\pi^{\mp}$, $\omega K^{+}K^{-}\pi^{0}$, $\phi K^{0}_{S}K^{\pm}
\pi^{\mp}$, and $\phi K^{+}K^{-}\pi^{0}$ are
$2(\pi^{+}\pi^{-})K^{\pm}\pi^{\mp}\gamma\gamma$,
$\pi^{+}\pi^{-}K^{+}K^{-}\gamma \gamma \gamma \gamma$,
$K^{+}K^{-}\pi^{+}\pi^{-}K^{\pm}\pi^{\mp}$, and $2(K^{+}K^{-})\gamma
\gamma$, respectively.\\ \indent Candidate events are required to
satisfy the following common selection criteria:
\begin{enumerate}
\item The correct number of charged tracks with net charge zero is
required for each event. Each charged track should have a good helix
fit in the MDC, and the polar angle $\theta$ of each track in the MDC
must satisfy $|\cos \theta|<0.8$. The event must originate from the
collision point; tracks except $\pi^{\pm}$ from $K^{0}_{S}$ must
satisfy $\sqrt{x^{2}+y^{2}} \leq 2$ cm, $|z| \leq 20$ cm, where $x$,
$y$, and $z$ are the space coordinates of the point of closest
approach of tracks to the beam axis.

\item Candidate events should have at least the minimum number of
 isolated photons associated with the different final states. Each
 photon should have an energy deposit in the BSC greater than $50$ MeV, the
 angle between the shower development direction and the photon emission
 direction less than $30^{\circ}$, and the angle between the photon
 and any charged track larger than $8^{\circ}$.
 \item For each charged track in an event, $\chi^{2}_{PID}(i)$ is
 determined using both $dE/dx$ and TOF information:
$$\chi^{2}_{PID}(i) =\chi^{2}_{dE/dx}(i)+ \chi^{2}_{TOF}(i),$$ where
$i$ corresponds to the particle hypothesis. A charged track is
identified as a $K$ ($\pi$) if $\chi^{2}_{PID}$ for the $K$ ($\pi$)
hypothesis is less than those for the $\pi$ and $p$ ($K$ and $p$)
hypotheses.
\item The selected events are subjected to four constraint kinematic
fits (4C-fit), unless otherwise specified.  When there are more than
the minimum number of photons in an event, all combinations are tried,
and the combination with the smallest $\chi^{2}$ is retained.
\end{enumerate}
The branching fraction is calculated using
  \begin{eqnarray}
  B(J/\psi \rightarrow X )=\frac{N_{obs}}
  {\epsilon_{J/\psi \rightarrow X \rightarrow Y}
  \times N_{J/\psi} \times B(X \rightarrow Y)},
  \label{equ:all-branching-ratio}
  \end{eqnarray}
and the upper limit for a branching fraction
is determined using
\begin{eqnarray}
B(J/\psi \rightarrow X)
< \frac{N_{up}}{ \epsilon_{J/\psi \rightarrow X \rightarrow Y} \times N_{J/\psi} \times B(X \rightarrow Y)
\times (1-\sigma^{sys})},
\label{equ:upper-branching}
\end{eqnarray}
where, $N_{obs}$ is the number of events observed, $N_{up}$ is the
upper limit on the number of the observed events at the $90\%$
C.L. calculated using a Bayesian method ~\cite{pdg2006}, $\epsilon$ is the
detection efficiency obtained from MC simulation, $N_{J/\psi}$ is the
number of $J/\psi$ events, $(5.77 \pm 0.27) \times 10^{7}$
~\cite{fangss2003}, $\sigma^{sys}$ is the corresponding systematic
error, and $B(X \rightarrow Y)$ is the branching fraction, taken from
the Particle Data Group (PDG) ~\cite{pdg2006}, of the $X$ intermediate
state to the $Y$ final state.

\subsection{\bf $J/\psi \rightarrow \{\omega, \eta \} K^{0}_{S}K^{\pm}\pi^{\mp}$}

\begin{figure}[htbp]
  \centering
\includegraphics[width=0.35\textwidth]{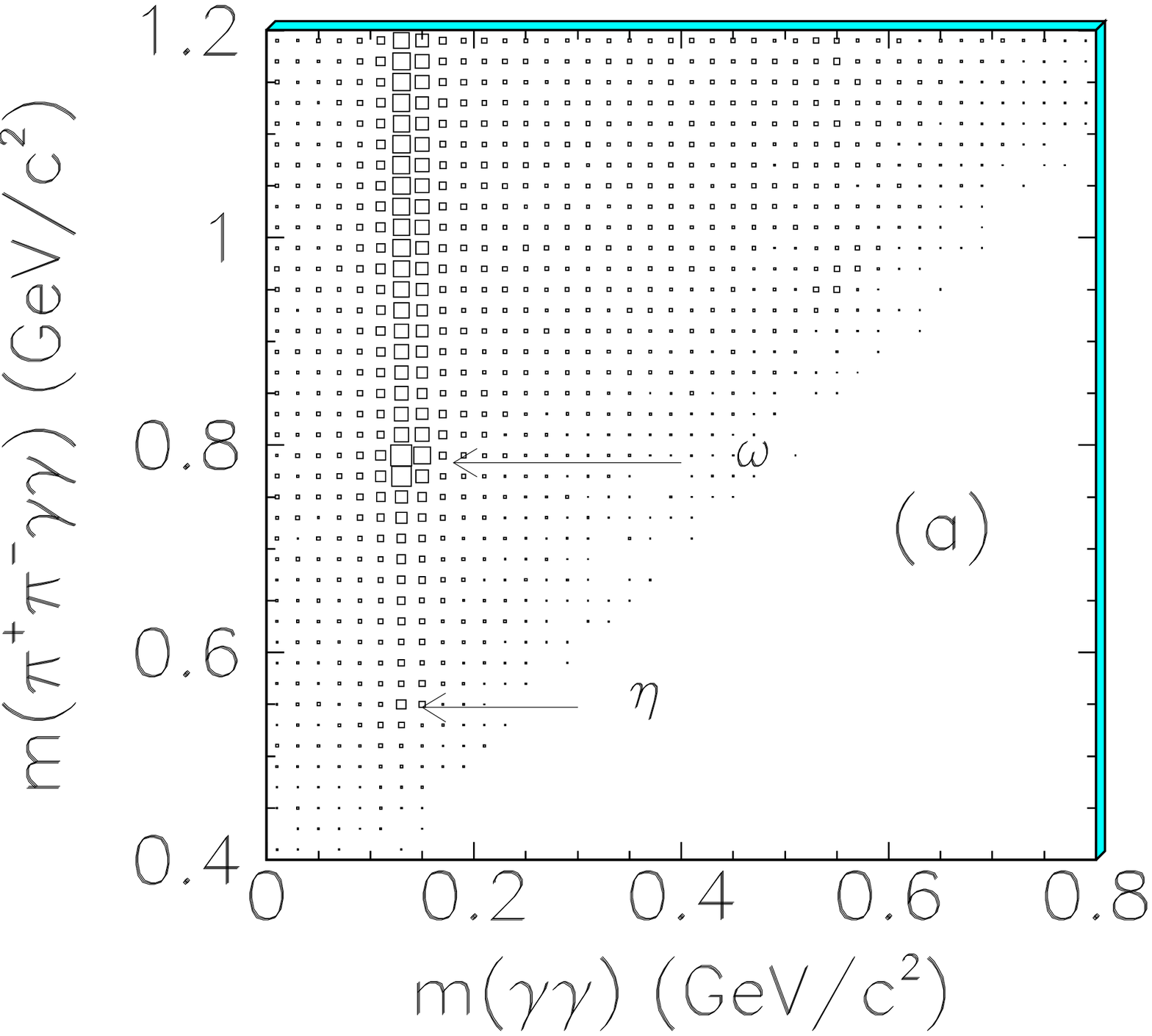}
\includegraphics[width=0.35\textwidth]{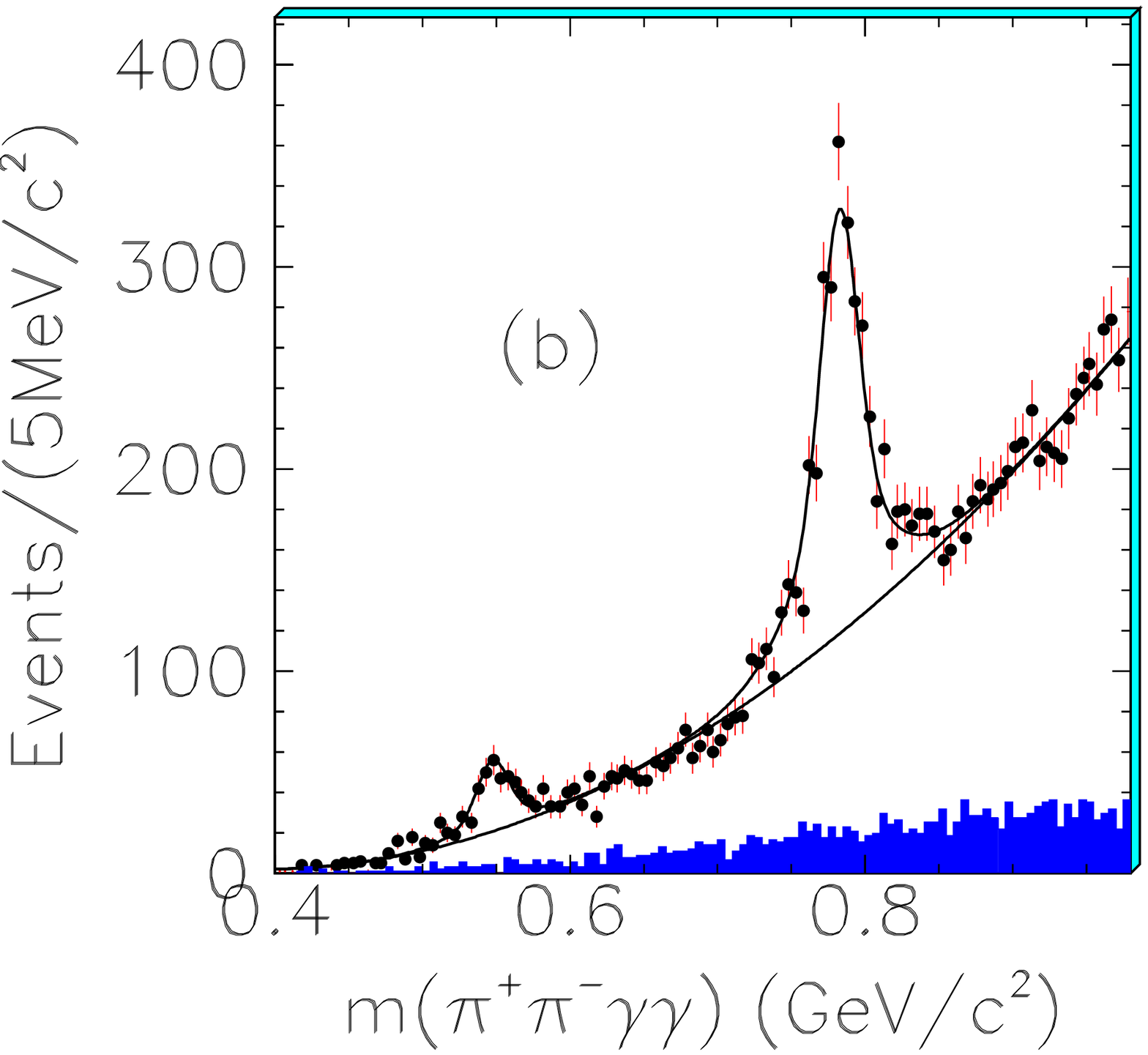}
  \caption{(a) The scatter plot of $m_{\gamma \gamma}$ versus
  $m_{\pi^{+}\pi^{-}\gamma \gamma}$, and (b) the $\pi^{+}\pi^{-}
  \gamma \gamma$ invariant mass for $J/\psi \rightarrow \{\omega,
  \eta\}K^{0}_{S}K^{\pm}\pi^{\mp}$ candidate events with two possible
  entries per event.  The curves in (b) are the results of the fit described
  in the text, and the shaded histogram in (b) shows the normalized
  background estimated from the $K^{0}_{S}$-sideband region ($0.025$
  GeV/$c^{2}<|m_{\pi^{+}\pi^{-}}-m_{K^{0}_{S}}|<0.055$ GeV/$c^{2}$).}
             \label{fig:w-pi0}
\end{figure}

At least one charged track must be identified as a kaon using TOF and
$dE/dx$ information.  If there is more than one kaon candidate, the
assigned kaon is the one with the largest kaon weight. Candidate
events are fitted kinematically using energy momentum conservation
(4C-fit) under the $2(\pi^{+}\pi^{-})K^{\pm}\pi^{\mp} \gamma \gamma$
hypothesis, and $\chi^{2}<25$ is required. Each event is required to
contain one $K^{0}_{S}$ meson with six possible
$\pi^{+}\pi^{-}$ combinations to test for consistency with the
$K^{0}_{S}$. Looping over all combinations, we select the one
closest to the $K^{0}_{S}$ mass, denoted as $m_{\pi^{+}\pi^{-}}$, provided it is
within $15$ MeV/$c^{2}$ of the $K^{0}_{S}$ mass.  After $K^{0}_{S}$
selection, the two remaining oppositely-charged pion combinations
along with the two gammas are used to calculate
$m_{\pi^{+}\pi^{-}\gamma \gamma}$. Figure~\ref{fig:w-pi0} (a) shows
the scatter plot of $m_{\gamma \gamma}$ versus
$m_{\pi^{+}\pi^{-}\gamma \gamma}$ with two possible entries per event,
where clear $\eta$ and $\omega$ signals are seen.

The $\pi^{+} \pi^{-} \gamma \gamma$ invariant mass distribution with
two possible entries per event is shown in Fig. \ref{fig:w-pi0} (b),
where $\eta$ and $\omega$ signals are apparent. The branching
fractions of $J/\psi \to \omega K^{0}_{S}K^{\pm}\pi^{\mp}$ and $\eta
K^{0}_{S}K^{\pm}\pi^{\mp}$ are obtained by fitting this
distribution. The backgrounds for $J/\psi \to \omega
K^{0}_{S}K^{\pm}\pi^{\mp}$ which contribute to the peak in the
$\omega$ signal region mainly come from non-$K^{0}_{S}$ events and
events from $J/\psi \rightarrow \omega K^{0}_{S} K^{0}_{S}$ that
survive selection criteria.  The number of background events from
$J/\psi \rightarrow \omega K^{0}_{S} K^{0}_{S}$ is estimated from
Monte-Carlo simulation to be less than 2 . Backgrounds for $J/\psi \to
\eta K^{0}_{S}K^{\pm}\pi^{\mp}$ contributing to the peak in the $\eta$
signal region mainly come from non-$K^{0}_{S}$ events and events from
$J/\psi$ decays into $K^{*0}\bar{K^{*}_{2}}(1430)^{0}\rightarrow
(K^{0}_{S} \pi^{0})(K^{0}_{S} \eta)$. The latter contribution is
estimated to be less than one event from MC simulation.
Non-$K^{0}_{S}$ events from the $K^{0}_{S}$ sideband region (0.025
GeV/c$^2$ $<|m_{\pi^+\pi^-}-m_{K^{0}_{S}}|<0.055$ GeV/c$^{2}$) are shown in Fig. \ref{fig:w-pi0} (b) as the shaded
histogram, the background events are
$19.2\pm 15.6$ $\omega$ and $-4.1\pm7.0$ $\eta$ by fitting the distribution with possible signals and polynomial background.

A fit to the $m_{\pi^{+} \pi^{-} \gamma \gamma}$ distribution is
performed by using $\omega$ and $\eta$ Breit-Wigner (BW) functions
folded with Gaussian resolution functions plus a quadratic polynomial,
shown as the curve in Fig. \ref{fig:w-pi0} (b). The numbers of events
in the $\omega$ and $\eta$ peaks are $1971.7 \pm 41.4$ and $231.6 \pm
23.1$, respectively. Here, the background events in the decays
of $J/\psi \to \omega K^{0}_{S}K^{\pm}\pi^{\mp}$ and $J/\psi
\rightarrow\eta K^{0}_{S}K^{\pm}\pi^{\mp}$ estimated above are not
subtracted but are included in the background systematic error. The
$J/\psi \rightarrow \omega K^{0}_{S} K^{\pm} \pi^{\mp}$ and $J/\psi
\rightarrow \eta K^{0}_{S} K^{\pm} \pi^{\mp}$ detection efficiencies
are obtained from MC simulation to be $1.48\%$ and $1.18\%$,
respectively. The branching fractions are then determined as:

\begin{eqnarray}
B(J/\psi \rightarrow \omega K^{0}_{S} K^{\pm} \pi^{\mp}) & =
&(3.77 \pm 0.08) \times 10^{-3},\nonumber
\end{eqnarray}
\begin{eqnarray}
B(J/\psi \rightarrow \eta K^{0}_{S} K^{\pm} \pi^{\mp})
& = & (2.18 \pm 0.22) \times 10^{-3}.\nonumber
\end{eqnarray}

Here the errors are statistical only.

\subsubsection{\bf $J/\psi \to \omega K^{*}\bar{K}+c.c. \rightarrow \omega K^{0}_{S}K^{\pm}\pi^{\mp}$}

\begin{figure}[htbp]
  \centering
\includegraphics[width=0.35\textwidth]{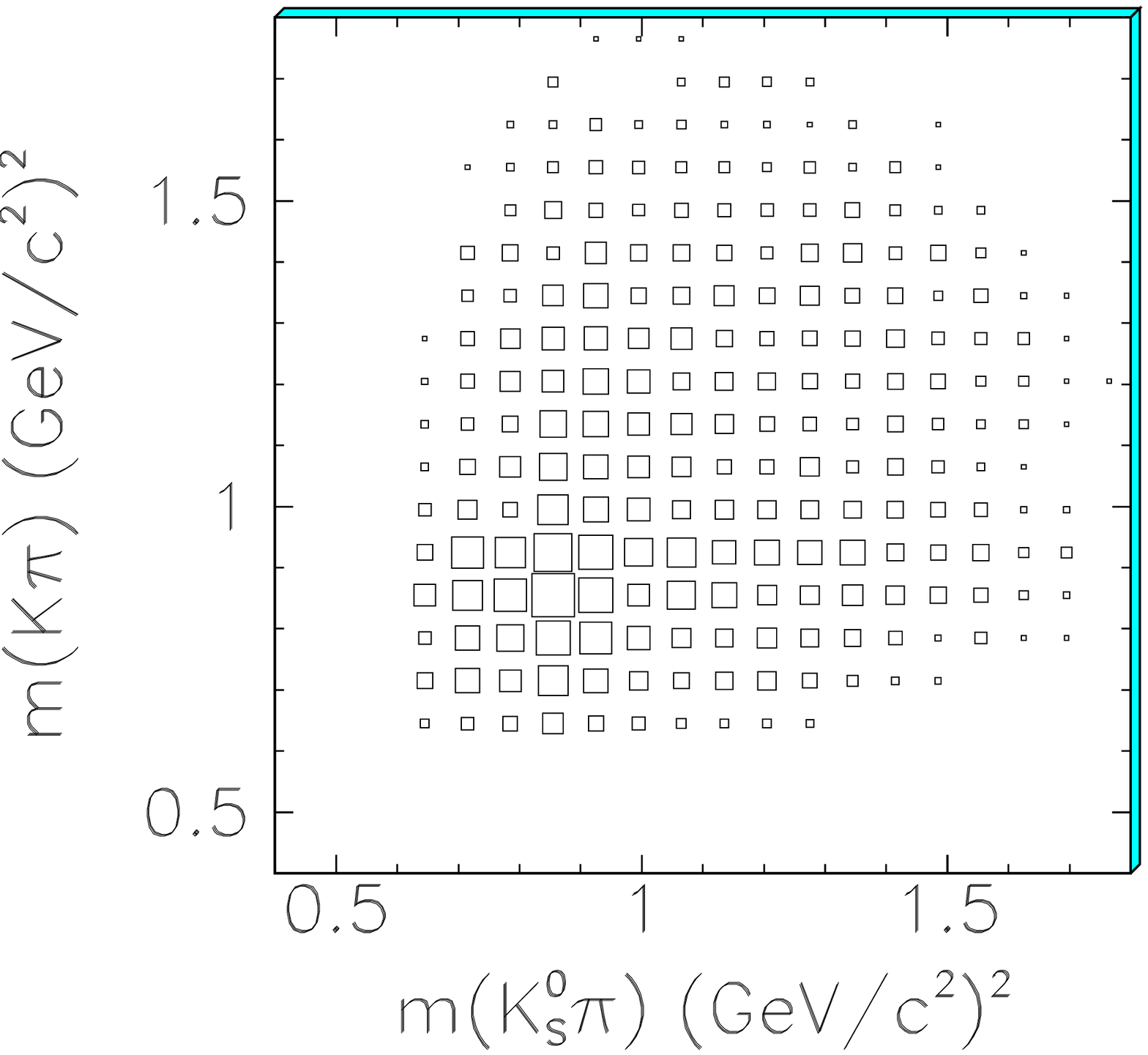}
  \caption{The scatter plot of $m_{K^{0}_{S}\pi}$ versus $m_{K^{\pm}\pi^{\mp}}$
  for $J/\psi \rightarrow \omega K^{0}_{S} K^{\pm} \pi^{\mp}$ candidate events.}
           \label{fig:ksp-kp-wkksp}
\end{figure}

To select the $\omega$ signal, the mass combination with $\pi^{+}
\pi^{-} \gamma \gamma$ closest to the $\omega$ mass is required to satisfy
$|m_{\pi^{+}\pi^{-}\gamma \gamma}-m_{\omega}|<0.04$ GeV/$c^{2}$.  Figure
\ref{fig:ksp-kp-wkksp} shows the scatter plot of
$m_{K^{0}_{S}\pi}$ versus $m_{K\pi}$ for $J/\psi \rightarrow \omega K^{0}_{S} K^{\pm} \pi^{\mp}$ candidate events,
where the events in the cross bands
correspond to $J/\psi \rightarrow \omega K^{*}\bar{K}+c.c.$.

\begin{figure}[htbp]
  \centering
\includegraphics[width=0.35\textwidth]{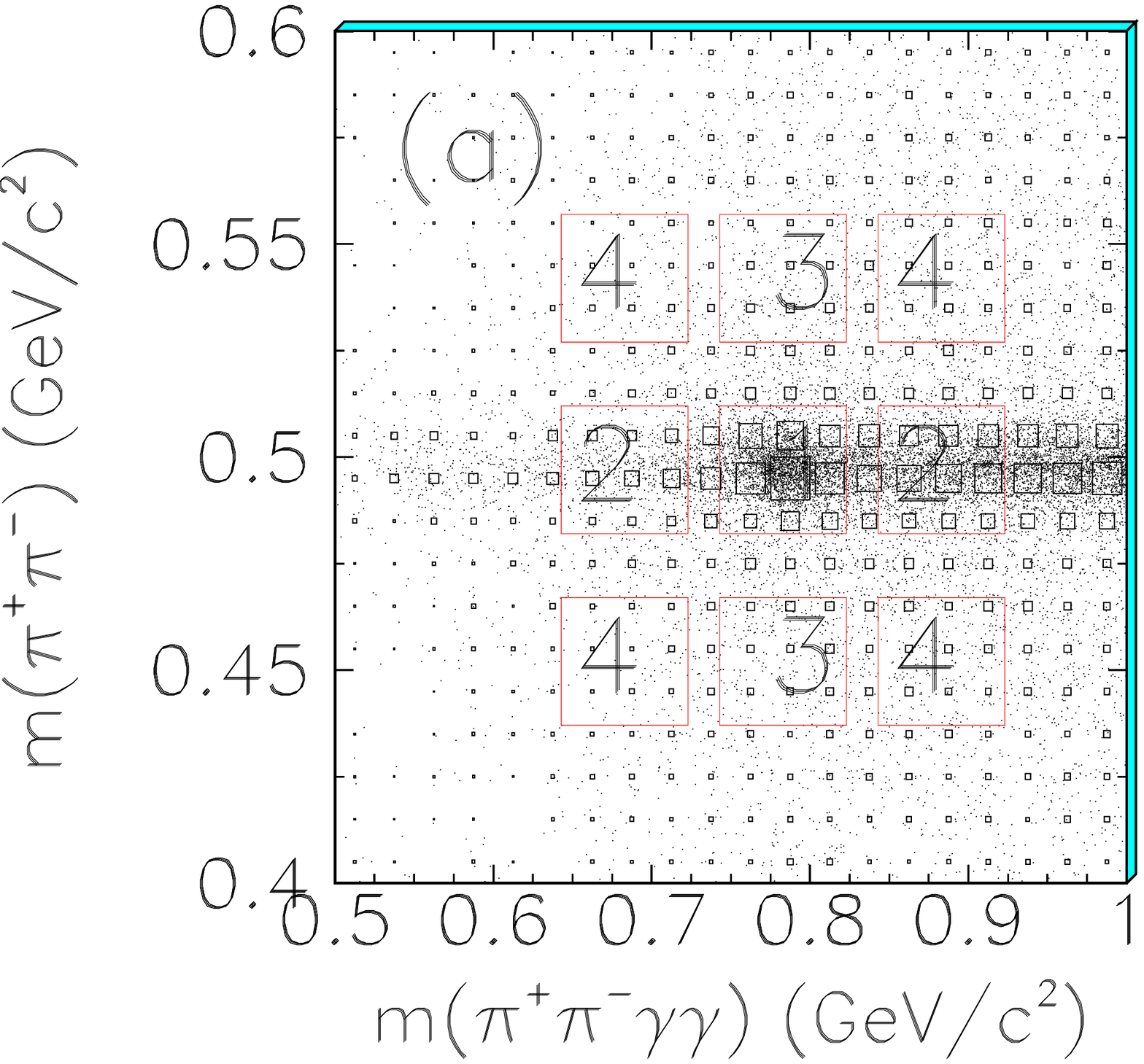}
\includegraphics[width=0.35\textwidth]{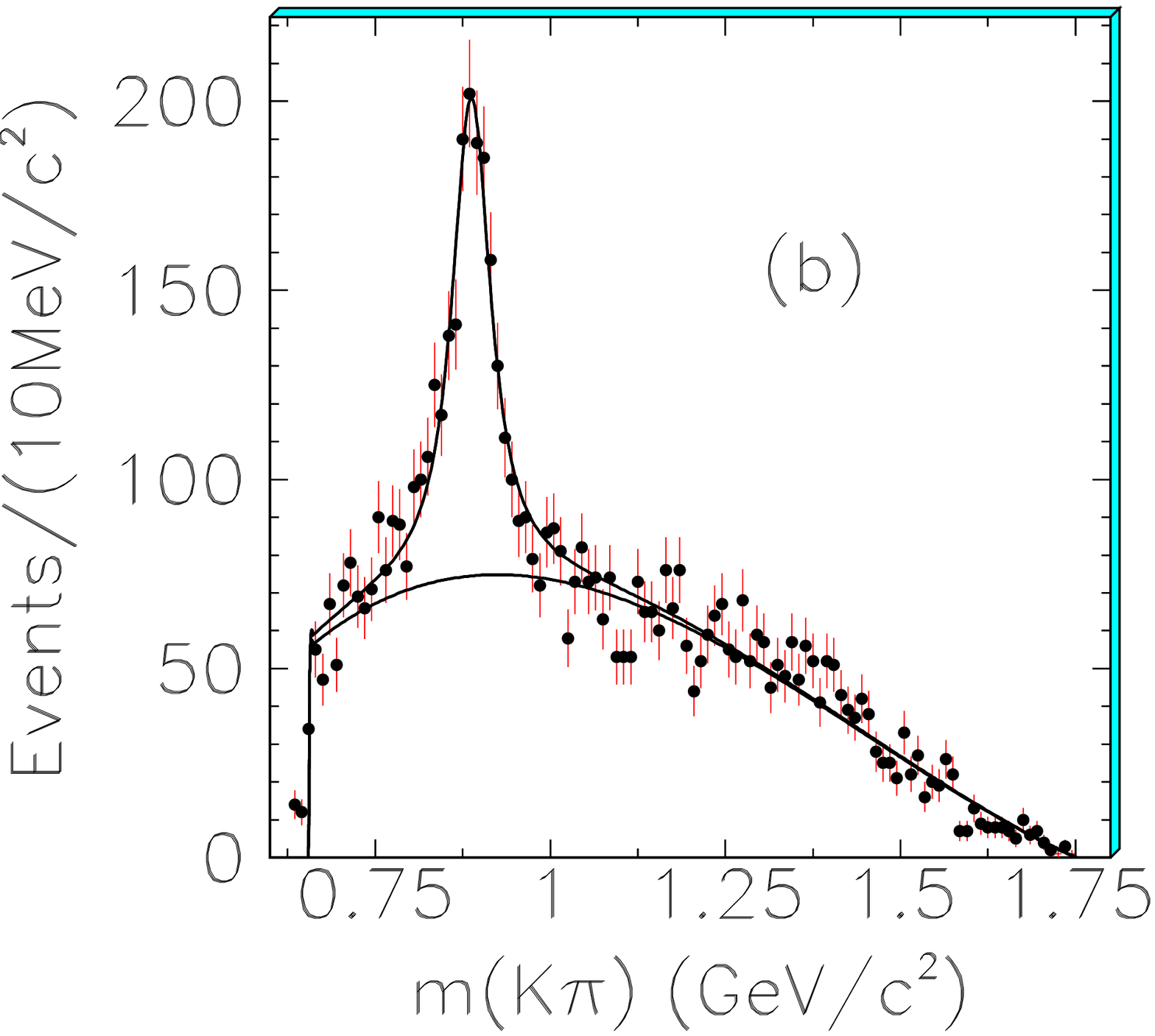}
  \centering
\includegraphics[width=0.35\textwidth]{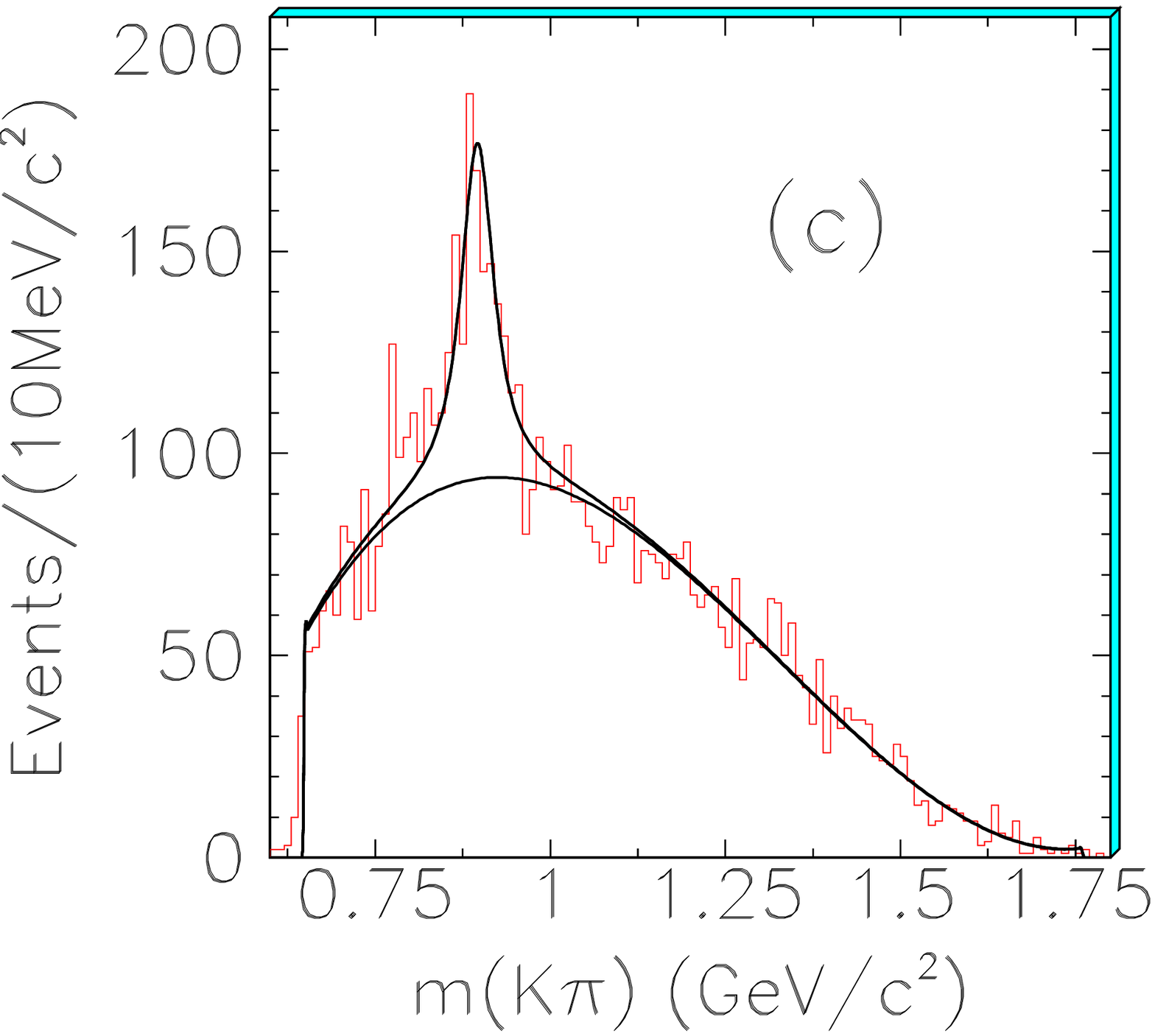}
\includegraphics[width=0.35\textwidth]{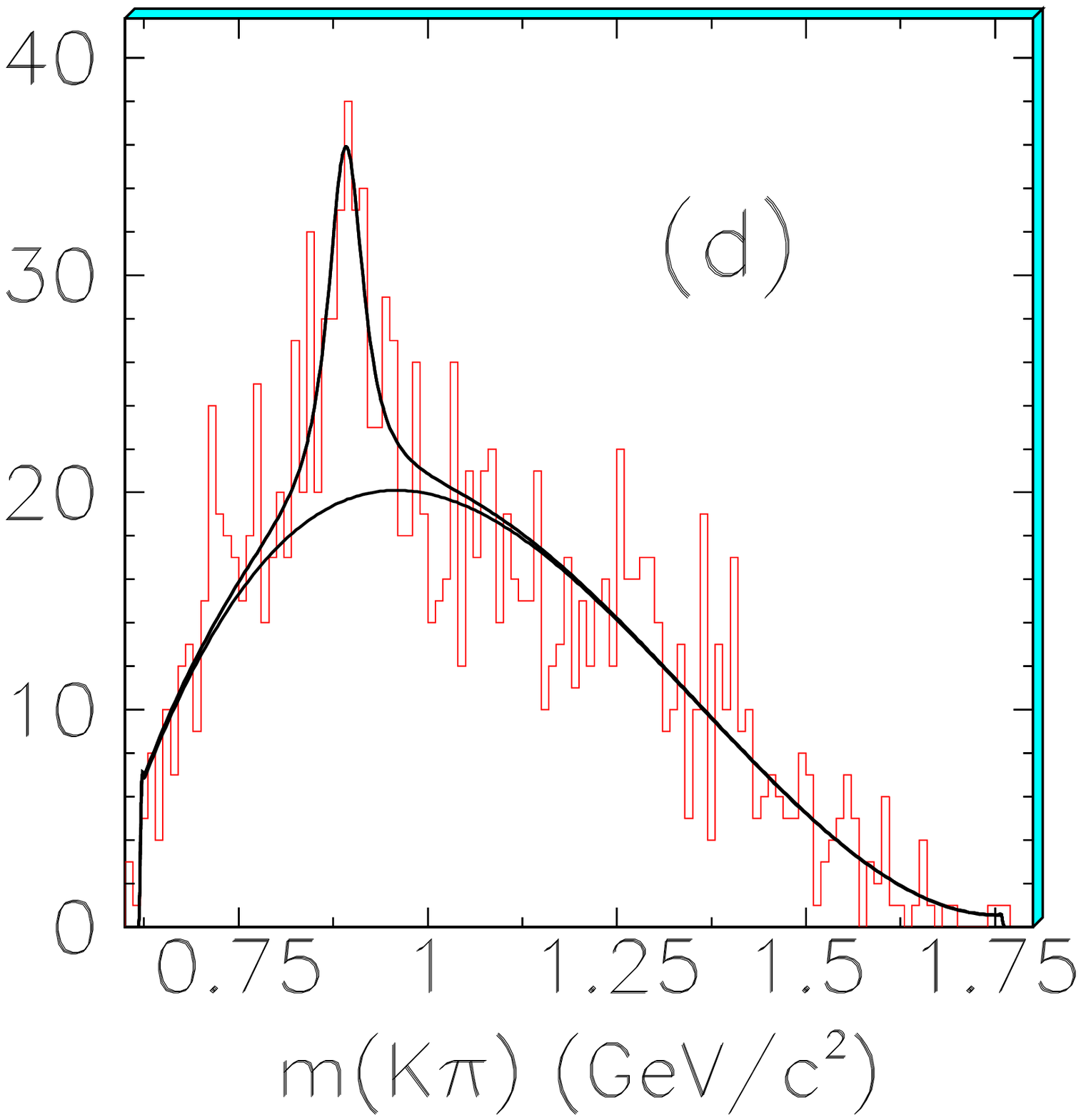}
  \caption{(a) The scatter plot of $m_{\pi^{+}\pi^{-}\gamma \gamma}$ versus
  $m_{\pi^{+}\pi^{-}}$, and the combined mass spectrum of $K^{0}_{S}
  \pi^{\mp}$ and $K^{\pm} \pi^{\mp}$ with two entries per event
  $J/\psi \rightarrow \omega K^{*}\bar{K}+c.c.$ candidate events for
  (b) the signal region (the central box 1); (c) the $\omega$ and
  $K^{0}_{S}$ sideband regions (two horizontal boxes 2 and two vertical
  sideband boxes 3); and for (d) the corner region (four diagonal
  boxes 4). The curves are the results of the fit described in the
  text.}
           \label{fig:k892-wkksp}
\end{figure}

Figure~\ref{fig:k892-wkksp} (a) shows the scatter plot of
$m_{\pi^{+}\pi^{-}\gamma \gamma}$ versus $m_{\pi^{+}\pi^{-}}$,
and there is an accumulation of events in the $\omega$ and $K^{0}_{S}$ cross bands.
The combined mass spectrum of $K^{0}_{S} \pi^{\mp}$ and $K^{\pm} \pi^{\mp}$ in the signal
region (box 1 in Fig.~\ref{fig:k892-wkksp} (a)),
which is defined as $|m_{\pi^{+}\pi^{-}}-m_{K^{0}_{S}}|<0.015$ GeV/$c^{2}$ and
$|m_{\pi^{+}\pi^{-}\gamma \gamma}-m_{\omega}|<0.04$ GeV/$c^{2}$,
is shown
in Fig. \ref{fig:k892-wkksp} (b), where a clear $K^{*}$ signal is observed. The $K^{*}$ signal is fitted
with a BW function folded with a Gaussian resolution function plus a third-order polynomial,
 and $1208.3\pm 93.3$ $K^{*}$ events are obtained.

Non-$\omega$ and non-$K^{0}_{S}$ backgrounds are studied using
$\omega$ and $K^{0}_{S}$ sideband events. Figure
\ref{fig:k892-wkksp} (c) is the fitted $K\pi$ mass spectrum in the
$\omega$ sideband region ($|m_{\pi^{+}\pi^{-}}-m_{K^{0}_{S}}|<0.015$
GeV/$c^{2}$, $0.06$
GeV$/c^{2}<|m_{\pi^{+}\pi^{-}\gamma \gamma}-m_{\omega}|<0.14$ GeV/$c^{2}$,
shown as horizontal sideband boxes $2$ in Fig.~\ref{fig:k892-wkksp} (a))
and $K^{0}_{S}$ sideband region ($0.03$
GeV/$c^{2}<|m_{\pi^{+}\pi^{-}}- m_{K^{0}_{S}}|<0.06$ GeV/$c^{2}$,
$|m_{\pi^{+}\pi^{-}\gamma \gamma}-m_{\omega}|<0.04$ GeV/$c^{2}$, shown as
vertical sideband boxes $3$), and the number of $K^{*}$ sideband events
$N_{sid1}=(686.2\pm 56.0)$ is obtained. Figure \ref{fig:k892-wkksp}
(d) is background from the corner region ($0.03$
GeV/$c^{2}<|m_{\pi^{+}\pi^{-}}-m_{K^{0}_{S}}|<0.06$ GeV/$c^{2}$,
$0.06$ GeV/$c^{2}<|m_{\pi^{+}\pi^{-}\gamma \gamma}-m_{\omega}|<0.14$
GeV/$c^{2}$, shown as diagonal boxes $4$), and the number of $K^{*}$ events
$N_{sid2}$ is equal to $(134.1 \pm 25.5)$. The number of background
events in the signal region is half of the sum of $K^{*}$ events
in the $\omega$ sideband and $K^{0}_{S}$ sideband regions
($N_{sid1}$) minus a quarter of the $K^{*}$ events in the corner
regions ($N_{sid2}$). So $N_{bg}=(686.2\pm 56.0)/2-(134.1 \pm
25.5)/4= (309.6\pm28.8)$.

The detection efficiency is estimated to be $1.23\%$ from MC simulation. After background
subtraction, the
branching fraction is determined to be
\begin{eqnarray}
B(J/\psi \rightarrow \omega K^{*} \bar{K}+c.c.) & = &(6.20 \pm 0.68)
\times 10^{-3},\nonumber
\end{eqnarray}
where the error is statistical only.

\subsubsection{\bf $J/\psi \to \omega X(1440)\rightarrow \omega K^{0}_{S}K^{\pm}\pi^{\mp}$}

\begin{figure}[htbp]
  \centering
\includegraphics[width=0.35\textwidth]{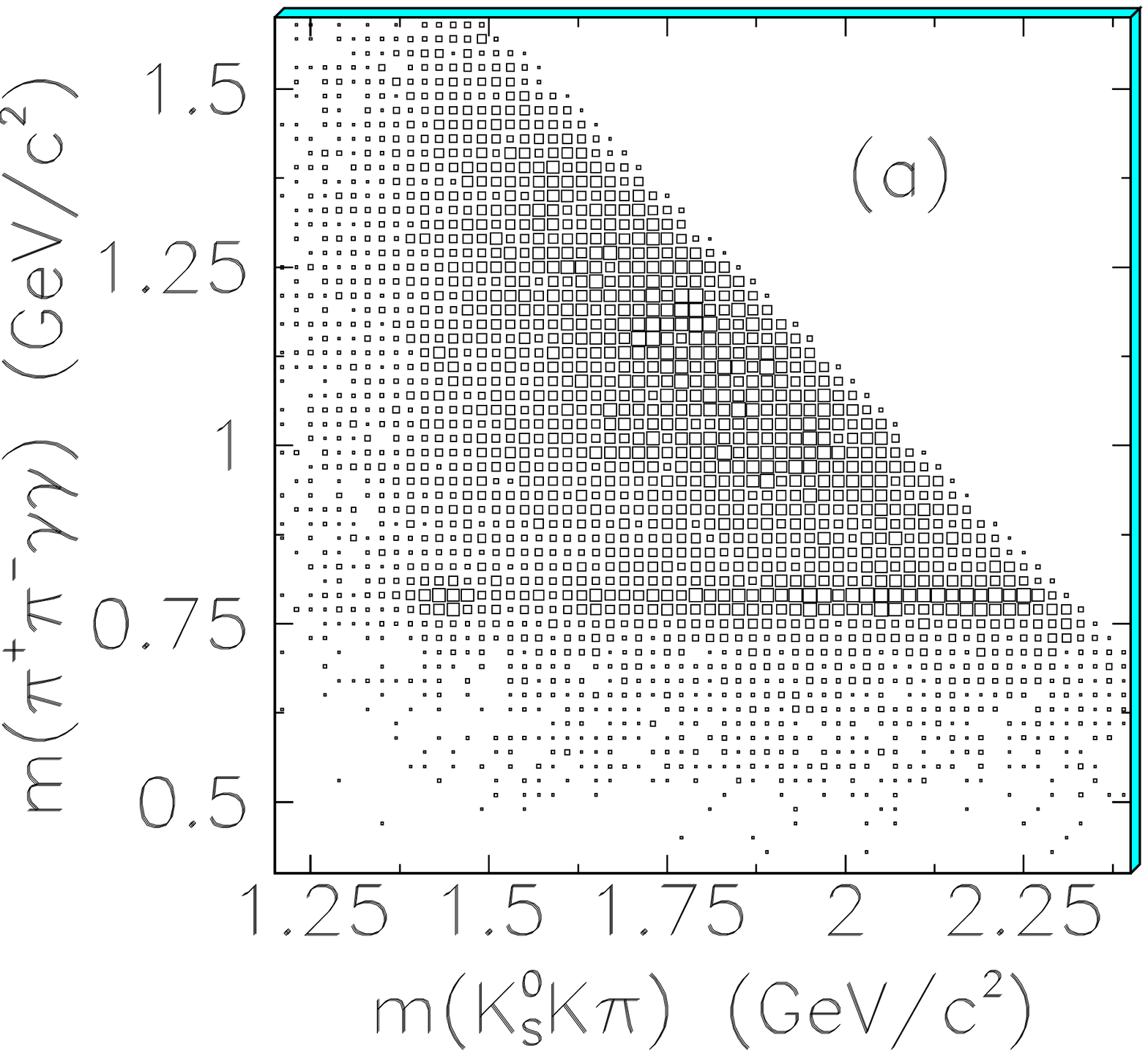}
\includegraphics[width=0.35\textwidth]{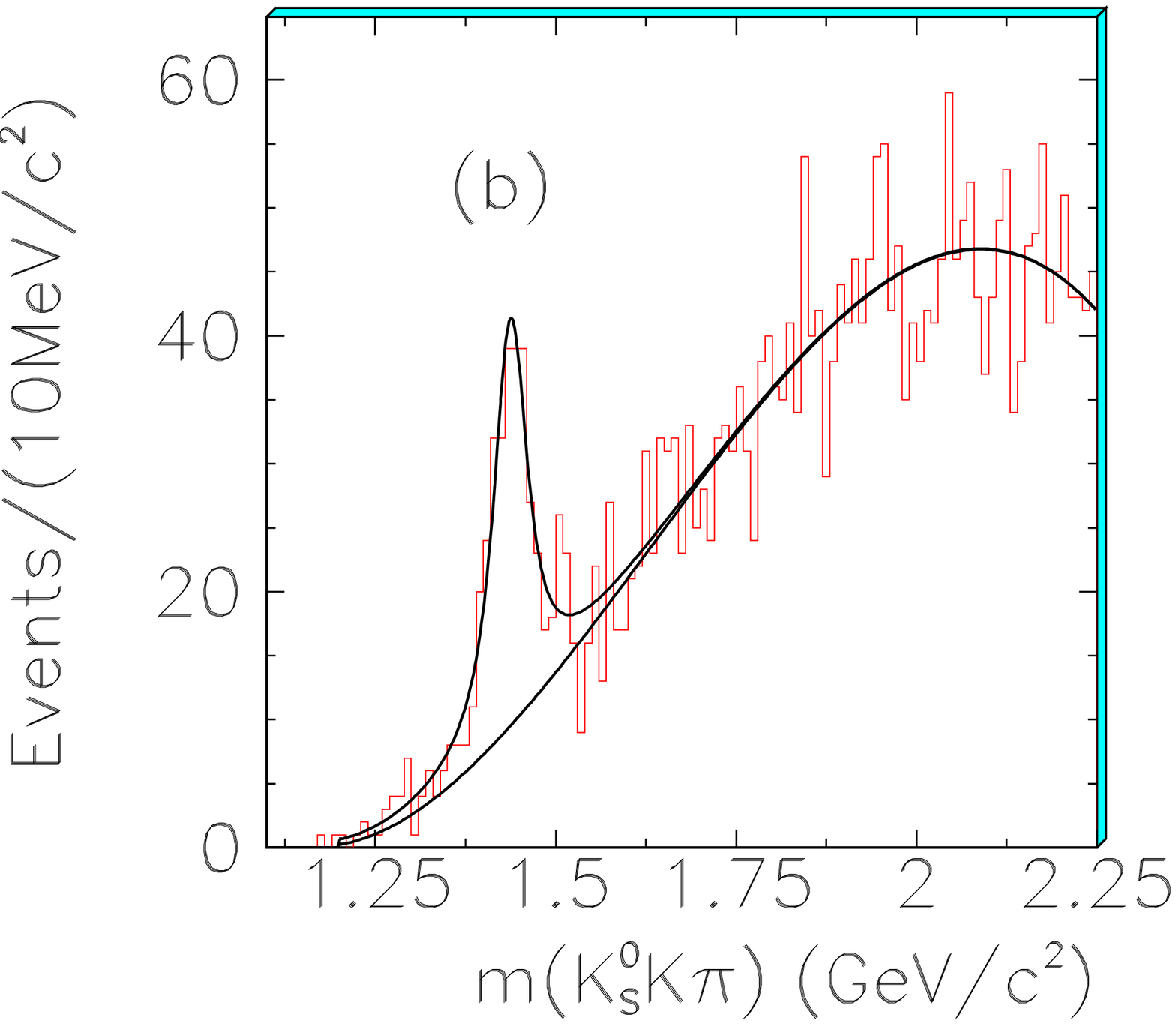}
  \caption{(a) The scatter plot of $m_{K^{0}_{S}K^{\pm}\pi^{\mp}}$
  versus $m_{\pi^{+}\pi^{-}\gamma \gamma}$ and (b) the $K^{0}_{S} K^{\pm}
  \pi^{\mp}$ invariant mass distribution for $J/\psi \rightarrow
  \omega K^{0}_{S}K^{\pm}\pi^{\mp}$ candidate events.  The curves in (b)
  are the results of the fit described in the text. }
           \label{fig:w-x1440-recoiling}
\end{figure}

Figure \ref{fig:w-x1440-recoiling} (a) shows the scatter plot of
$m_{K^{0}_{S}K^{\pm}\pi^{\mp}}$ versus $m_{\pi^{+}\pi^{-}\gamma \gamma}$,
and Fig. \ref{fig:w-x1440-recoiling} (b) is the
$K^{0}_{S}K^{\pm}\pi^{\mp}$ invariant mass spectrum after $\omega$
selection ($|m_{\pi^{+}\pi^{-}\gamma \gamma}-m_{\omega}|<0.04$
GeV/c$^{2}$). Figs. \ref{fig:w-x1440-recoiling} (a) and (b) show a
resonance near $1.44$ GeV/$c^{2}$, denoted as $X(1440)$. To ensure
that this peak is not due to background, we have made
studies of potential background processes using both data and MC
simulations. Non-$\omega$ and non-$K^{0}_{S}$ processes are studied
with $\omega$ and $K_S^0$ mass sideband events, respectively. The main
background channel $J/\psi \rightarrow \omega 2(\pi^{+}\pi^{-})$ and
other background processes with 6-prong events are studied by MC
simulation. In addition, we also checked for possible backgrounds with
a MC sample of $60 \times 10^{6} ~J/\psi \to anything$ decays
generated by the LUND-charm model~\cite{jcchen2004}. None of these
background sources produces a peak around $1.44$ GeV/$c^{2}$ in the
$K^{0}_{S}K^{\pm}\pi^{\mp}$ invariant mass spectrum.

The $K^{0}_{S}K^{\pm}\pi^{\mp}$ invariant mass distribution is fitted
with a BW function convoluted with a Gaussian mass resolution function
($\sigma=7.44$ MeV/$c^{2}$) to represent the $X(1440)$ signal and a
third-order polynomial background function. The mass and width
obtained from the fit are $M=1437.6 \pm 3.2$ MeV/$c^{2}$ and
$\Gamma=48.9 \pm 9.0$ MeV/$c^{2}$, and the fit yields $248.8
\pm 35.2$ events.

 Using the efficiency of $1.45\%$ determined from a uniform
 phase space MC simulation, we obtain the branching fraction to be
  \begin{eqnarray}
B(J/\psi \rightarrow \omega X(1440))\cdot B( X(1440) \rightarrow K^{0}_{S} K^{\pm} \pi^{\mp})
& = &(4.86 \pm 0.69) \times 10^{-4},\nonumber
\end{eqnarray}
where the error is only the statistical error.

\subsection{\bf $J/\psi \rightarrow \omega K^{+} K^{-} \pi^{0}$}

 At least one charged track is required to be a kaon and the combined
PID probability for $K^{+}K^{-} \pi^{+}\pi^{-}$ is required to be
greater than those for the $K^{\pm}\pi^{\mp} \pi^{+}\pi^{-}$ and
$\pi^{+}\pi^{-}\pi^{+}\pi^{-}$ hypotheses.  A 4C kinematic fit is made
under the $K^{+} K^{-} \pi^{+} \pi^{-} 4 \gamma$ hypothesis. There are three combinations to form two $\pi^{0}$'s, and
further a six-constraint kinematic fit (6C-fit) with the smallest $\chi^{2}_{6C}$ is made requiring two $\pi^{0}$'s from four photons.
Events with
$\chi^{2}_{4C}<50$ and $\chi^{2}_{6C}<50$ are selected. To reject the
possible multiple photon background events, $\chi^{2}_{4C}$ is
required to be less than those for the $K^{+} K^{-} \pi^{+} \pi^{-} 2
\gamma$, $K^{+} K^{-} \pi^{+} \pi^{-} 3 \gamma$, and $K^{+} K^{-}
\pi^{+} \pi^{-} 5 \gamma$ hypotheses.  Background events with
$K^{0}_{S}$ decays, such as $K^{*0}\bar{K^{*}_{2}}(1430)^{0}
\rightarrow K^{0}_{S} K^{\pm} \pi^{\mp} \{\pi^{0}, 2\pi^{0}\}$, and
$\gamma K^{*} \bar{K^{*}} \rightarrow \gamma K^{0}_{S}K^{\pm}
\pi^{\mp} \pi^{0}$, are eliminated by requiring
$|m_{\pi^{+}\pi^{-}}-m_{K^{0}_{S}}|>0.02$ GeV/$c^{2}$ in the
$\pi^{+}\pi^{-}$ invariant mass.

\begin{figure}[htbp]
  \centering
\includegraphics[width=0.3\textwidth]{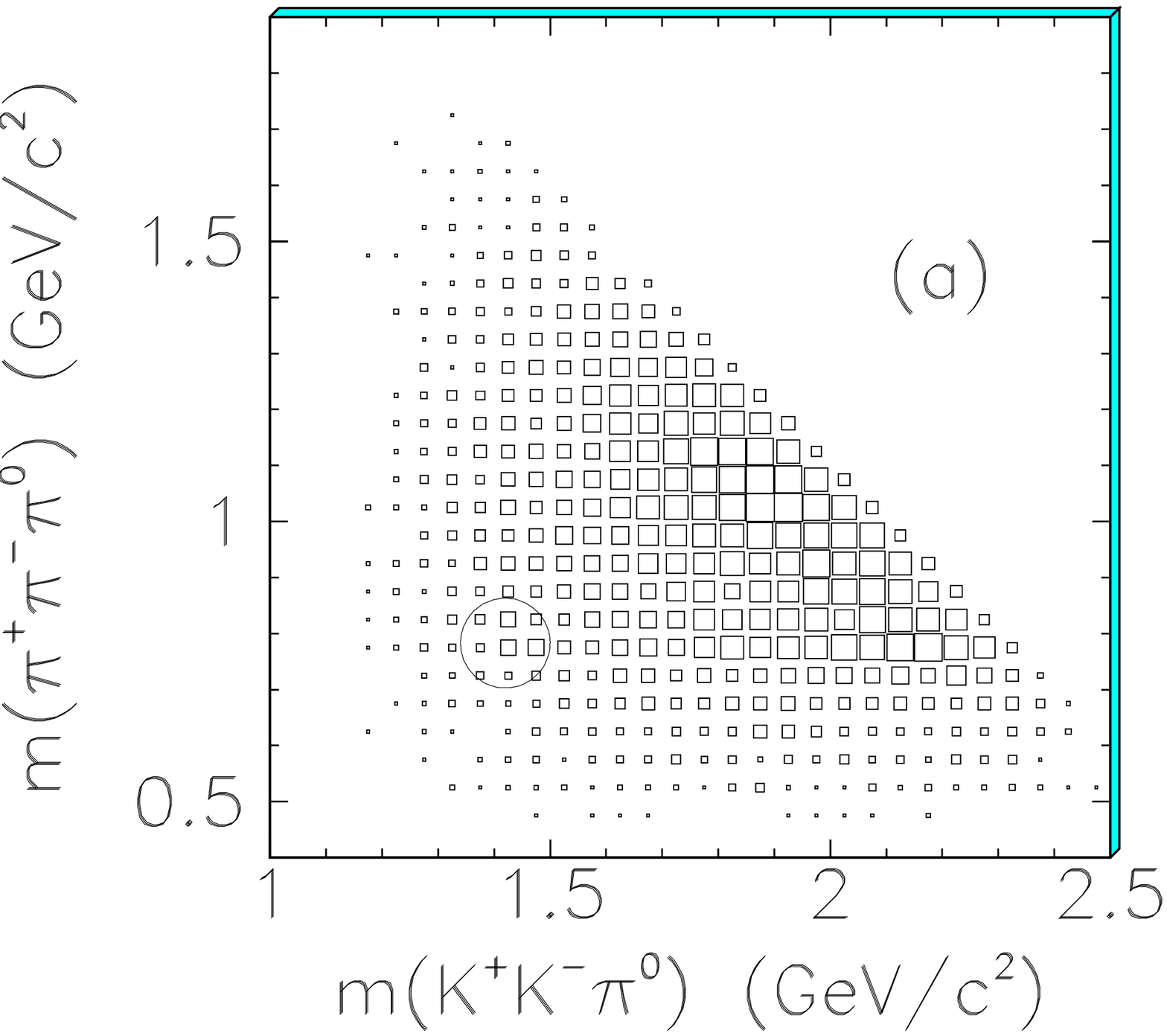}
\includegraphics[width=0.3\textwidth]{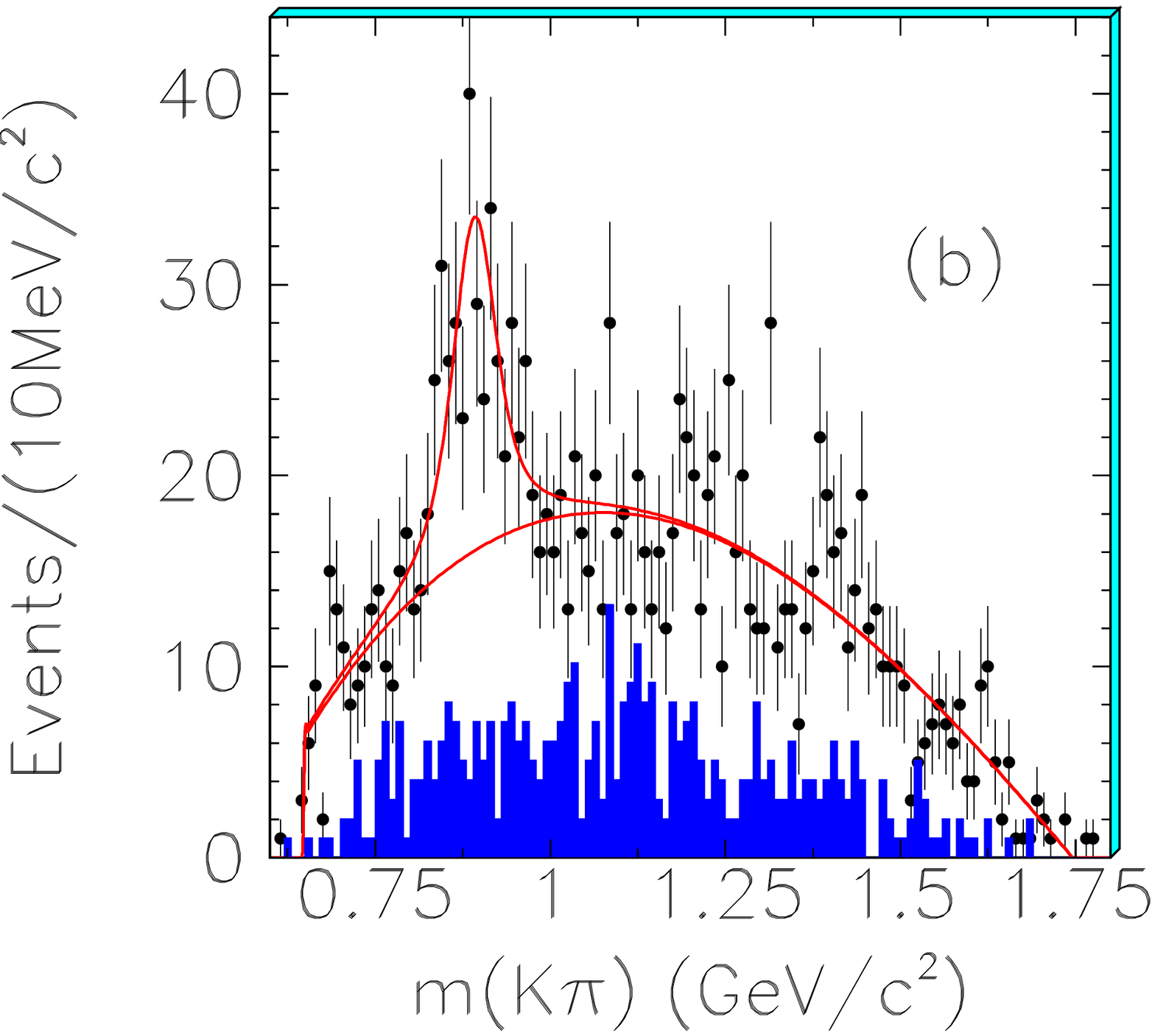}
\includegraphics[width=0.3\textwidth]{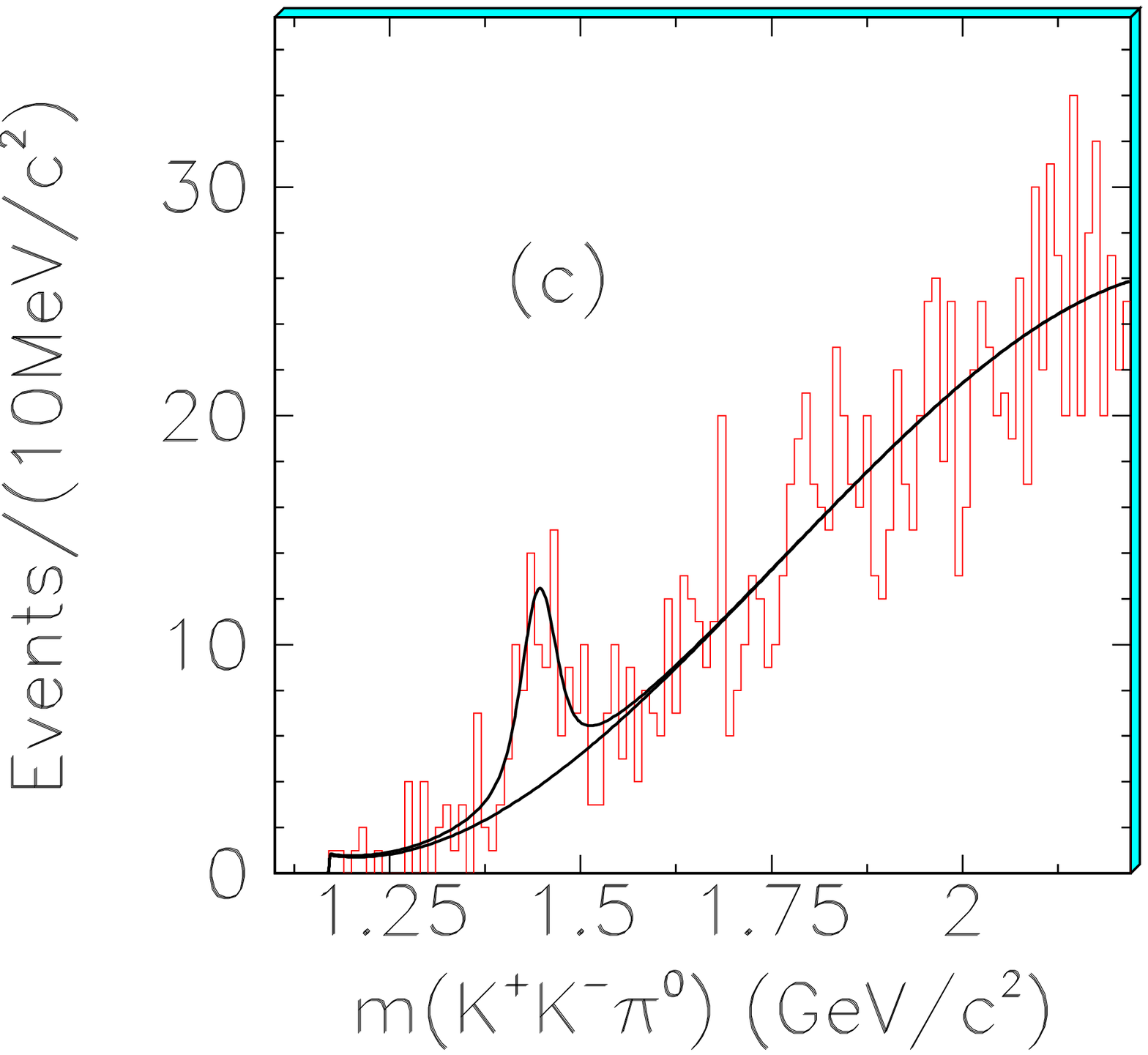}
  \caption{(a) The scatter plot of $m_{K^{+}K^{-}\pi^{0}}$ versus
$m_{\pi^{+}\pi^{-}\pi^{0}}$, (b) the $K^{\pm}\pi^{0}$ invariant mass
distribution with two possible entries per event, and (c) the $K^{+}
K^{-}\pi^{0}$ invariant mass distribution for $J/\psi \rightarrow
\pi^{+}\pi^{-}\pi^{0} K^{+}K^{-}\pi^{0}$ candidate events. The curves
are the results of the fit described in the text, and the shaded
histogram (b) shows the normalized background estimated from the
$\omega$-sideband region.}
          \label{fig:w-x1440-wkkp0}
\end{figure}

There are two $\pi^{+}\pi^{-}\pi^{0}$ mass combinations, and the one
closest to the $\omega$ mass, denoted as $m_{\pi^{+}\pi^{-}\pi^{0}}$,
is selected.  The scatter plot of $m_{K^{+}K^{-}\pi^{0}}$ versus
$m_{\pi^{+}\pi^{-}\pi^{0}}$ is shown in Fig.~\ref{fig:w-x1440-wkkp0}
(a), where the circle indicates some enhancement from $J/\psi \rightarrow \omega
X(1440)$ events in the $\omega K^{+}K^{-}\pi^{0}$ decay.

\subsubsection{\bf $J/\psi \to \omega K^{*\pm}K^\mp \rightarrow \omega K^{+}K^{-}\pi^{0}$}

To suppress the main $K^{*0}$ backgrounds,
$|m_{K^{\pm} \pi^{\mp}}-m_{K^{*0}}|>0.05$ GeV/$c^{2}$ is required.
In addition to the above selection, the further requirement of
$|m_{\pi^{+}\pi^{-}\pi^{0}}-m_{\omega}|<0.04$ GeV/$c^{2}$ is
imposed.  The combined
mass spectrum of $K^{+}\pi^{0}$ and $K^{-}\pi^{0}$ is shown in
Fig. \ref{fig:w-x1440-wkkp0} (b), where the $K^{*\pm}$ signal is seen,
and is fitted to obtain the branching fraction of $J/\psi \rightarrow
\omega K^{*\pm}K^{\mp}$.

Background events for $\omega K^{*\pm}K^{\mp}$ which could contribute to the
peak in the $K^{*\pm}$ signal region mainly come from events with
$K^{*}$ decays, such as $J/\psi \rightarrow K^{*0}
\bar{K^{*}_{2}}(1430)^{0}$ into 4-prong plus multiple photons, $J/\psi
\rightarrow \phi K^{*} \bar{K}$, and $J/\psi \rightarrow \gamma K^{*}
\bar{K^{*}}$, but their contributions can be ignored according to MC
studies. It is further confirmed that the $J/\psi \rightarrow \omega
K^{*\pm}K^{\mp}$ background is negligible using $\omega$ and
$\pi^{0}$ sideband events.

The $K^{\pm} \pi^{0}$ invariant mass distribution in
Fig. \ref{fig:w-x1440-wkkp0} (b) (2 entries/event) is fitted with a
$K^{* \pm}$ BW function with the mass and width fixed to
PDG values~\cite{pdg2006} plus a third-order polynomial. The
number of $K^{*\pm}$ events obtained is $(175.6\pm27.4)$. The
detection efficiency is $0.32\%$, and the branching fraction of $J/\psi
\rightarrow \omega K^{*} \bar{K}+c.c.$ is determined to be

\begin{eqnarray}
B(J/\psi \rightarrow \omega K^{*}\bar{K}+c.c.) & =
&(6.53\pm1.02) \times 10^{-3},\nonumber
\end{eqnarray}
where the error is statistical only.

\subsubsection{\bf $J/\psi \rightarrow \omega X(1440) \rightarrow \omega K^{+}K^{-}\pi^{0}$}

Figure \ref{fig:w-x1440-wkkp0} (c) shows the $K^{+}K^{-}\pi^{0}$
invariant mass recoiling against the $\omega$, where a $X(1440)$
signal is observed.  We have also studied potential
background processes using both data and MC simulations. Non-$\omega$
processes are studied with the $\omega$ mass sideband events ($0.06$
GeV/$c^{2}<|m_{\pi^{+}\pi^{-}\pi^{0}}-m_{\omega}| <0.10$
GeV/$c^{2}$). Background with $\omega$ decays is studied by MC
simulations, similar to those of $J/\psi \rightarrow \omega
K^{*}\bar{K}+c.c.  \rightarrow \omega K^{+}K^{-} \pi^{0}$. In
addition, we also checked for possible backgrounds using a MC sample
of $60 \times 10^{6} ~J/\psi \to anything$ decays generated by the
LUND-charm model. In each case, the $K^{+}K^{-}\pi^{0}$ mass
distribution shows no evidence of an enhancement near $1440$
MeV/$c^{2}$.

By fitting the $K^{+} K^{-}\pi^{0}$ mass spectrum in
Fig. \ref{fig:w-x1440-wkkp0} (c) with a BW function convoluted with a
Gaussian mass resolution function ($\sigma=14.2$ MeV/$c^{2}$) plus a
third-order polynomial background function, the mass and width of
$M=1445.9\pm 5.7$ MeV/$c^{2}$ and $\Gamma=34.2 \pm18.5$ MeV/$c^{2}$
are obtained, and the number of events from the fit is $62.1\pm18.3$. A
fit without a BW signal function returns a value of $-2 \ln L$ larger
than the nominal fit by 31.7 with three degrees of freedom (d.o.f.),
corresponding to a statistical significance of $5.0~\sigma$ for the
signal.

The efficiency is determined to be $0.64\%$ from a phase space MC simulation,
and the branching fraction is \\
\begin{eqnarray}
B(J/\psi \rightarrow \omega X(1440)) \cdot B(X(1440) \rightarrow K^{+} K^{-} \pi^{0})
& = &(1.92 \pm 0.57) \times 10^{-4},\nonumber
\end{eqnarray}
where the error is statistical.

\subsection{\bf $J/\psi \rightarrow \phi K^{0}_{S} K^{\pm} \pi^{\mp}$}

Events with six charged tracks are selected, and at least two charged
tracks must be identified as kaons. If there are more than two kaons,
the two kaons with the largest kaon PID probabilities are regarded as
the real kaons.  The other charged tracks are assumed, one at a time,
to be a kaon, while the other three to be pions, and these
combinations of three kaons and three pions are kinematically
fitted. The hypothesis with the smallest $\chi^{2}$ is considered as
the right combination, and $\chi^{2}<20$ is required.  Two
combinations of oppositely charged pions are used to reconstruct the
$K^{0}_{S}$ signal, and the one closest to the $K^{0}_{S}$ mass is
required to be within $15$ MeV/$c^{2}$.

\begin{figure}[htbp]
  \centering
\includegraphics[width=0.35\textwidth]{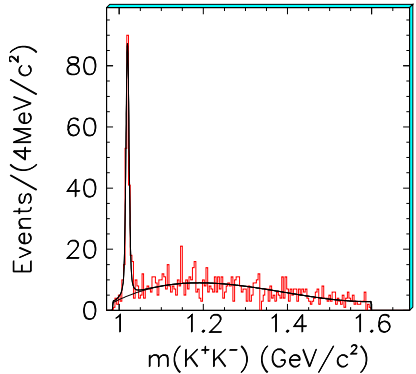}
  \centering
  \caption{The $K^{+}K^{-}$ invariant mass distribution for $J/\psi
  \rightarrow K^{+}K^{-} \pi^{+}\pi^{-} K^{\pm} \pi^{\mp}$ candidate
  events with two possible entries per event. The curves are the
  results of the fit described in the text.}
           \label{fig:phi-phikksp}
\end{figure}

The invariant mass of the two mass combinations formed with oppositely
charged kaons are shown in Fig. \ref{fig:phi-phikksp}, where a clear
$\phi$ signal is observed. A fit to the $K^{+}K^{-}$ mass distribution
in Fig \ref{fig:phi-phikksp} is performed to obtain the number of
$J/\psi \rightarrow\phi K^{0}_{S}K^{\pm}\pi^{\mp}$ events. Backgrounds
contributing to the $\phi$ signal peak mainly come from $J/\psi$ into
$\phi f^{\prime}_{2}(1525) \rightarrow \phi \eta \eta$, $\phi
\eta^{\prime} \rightarrow \phi \eta \pi^{+} \pi^{-}$, $\phi
K^{0}_{S}K^{0}_{S}$, and $\phi 2(\pi^{+}\pi^{-})$ (excluding $\phi
K^{0}_{S}K^{0}_{S}$).  From MC simulations of these background
channels, the number of background $\phi$ events in the signal region
is less than one, and $K^{0}_{S}$-sideband events also show that
the background is negligible.

The $K^{+} K^{-}$ mass distribution in Fig. \ref{fig:phi-phikksp} is
fitted with a BW function convoluted with a Gaussian mass resolution
function ($\sigma=2.93$ MeV/$c^{2}$) plus a third-order polynomial
background function.  The number of $\phi$ events from the fit is
$227.1 \pm 19.0$. Using the detection efficiency of $1.56\%$, the
corresponding branching fraction is

\begin{eqnarray}
B(J/\psi \rightarrow \phi K^{0}_{S}K^{\pm}\pi^{\mp})
& = &(7.37 \pm 0.62) \times 10^{-4},\nonumber
\end{eqnarray}
where the error is statistical only.

\subsubsection{\bf $J/\psi \to \phi K^{*} \bar{K}+c.c. \rightarrow \phi K^{0}_{S} K^{\pm} \pi^{\mp}$}

\begin{figure}[htbp]
  \centering
\includegraphics[width=0.3\textwidth]{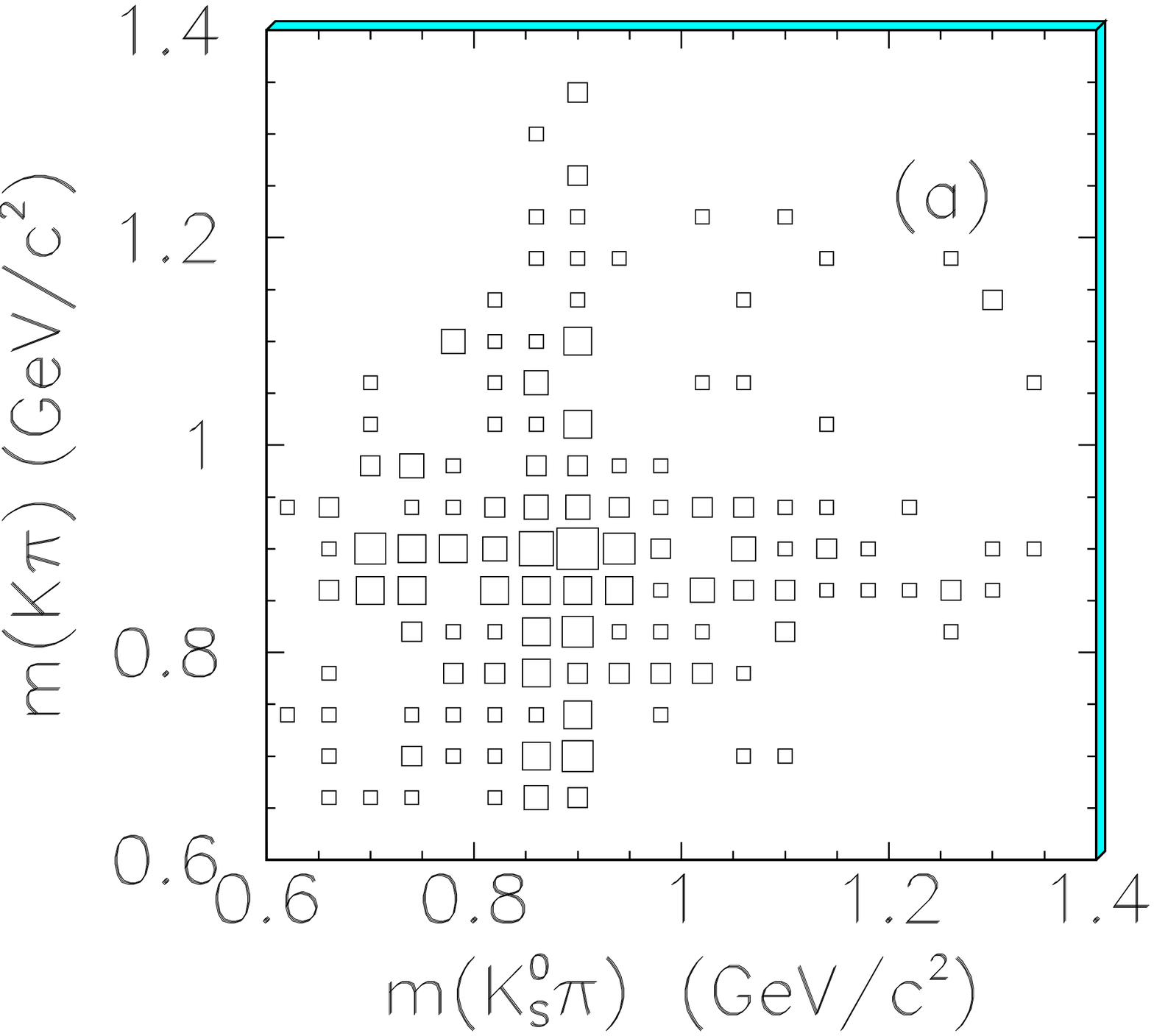}
\includegraphics[width=0.3\textwidth]{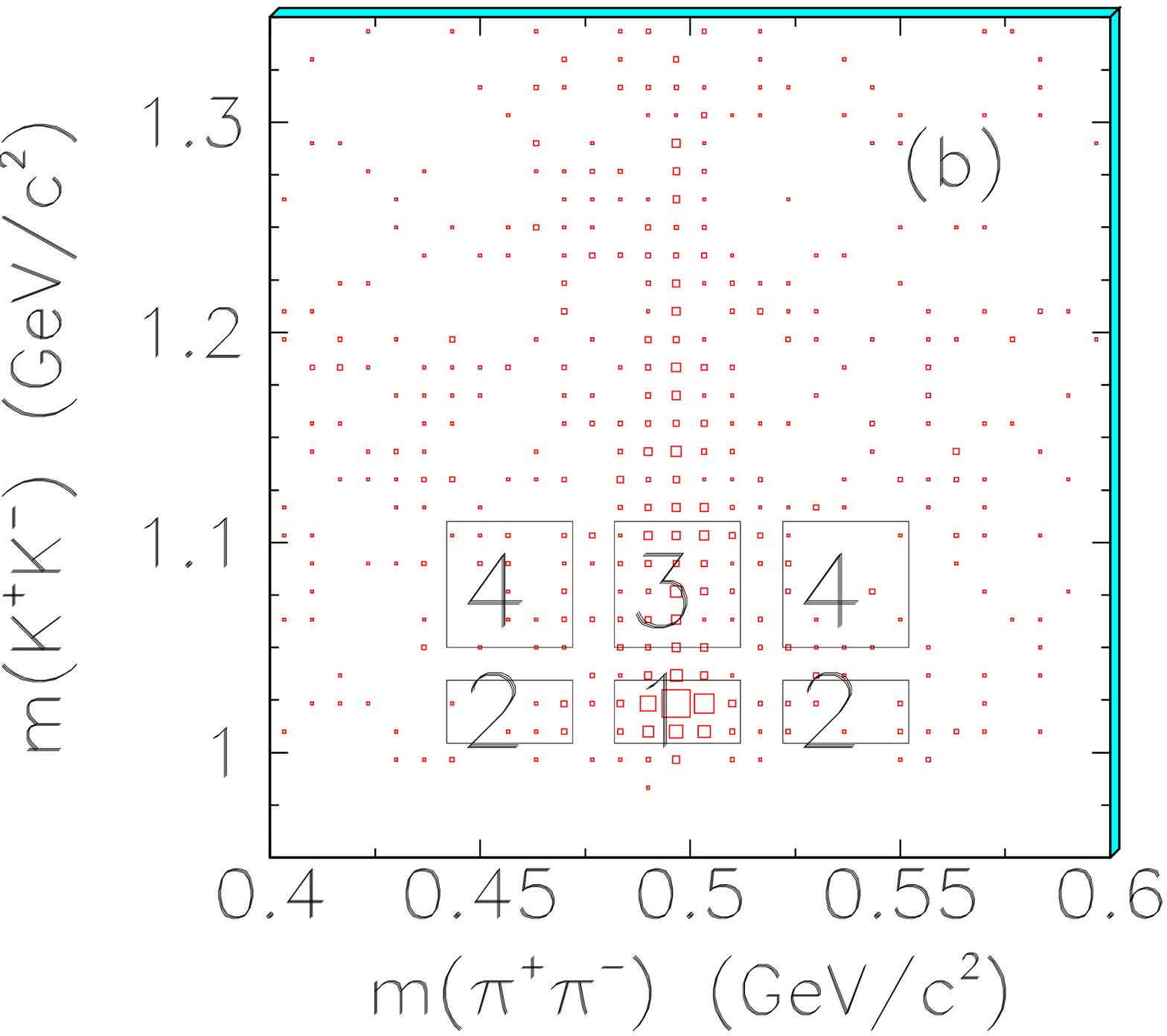}
\includegraphics[width=0.3\textwidth]{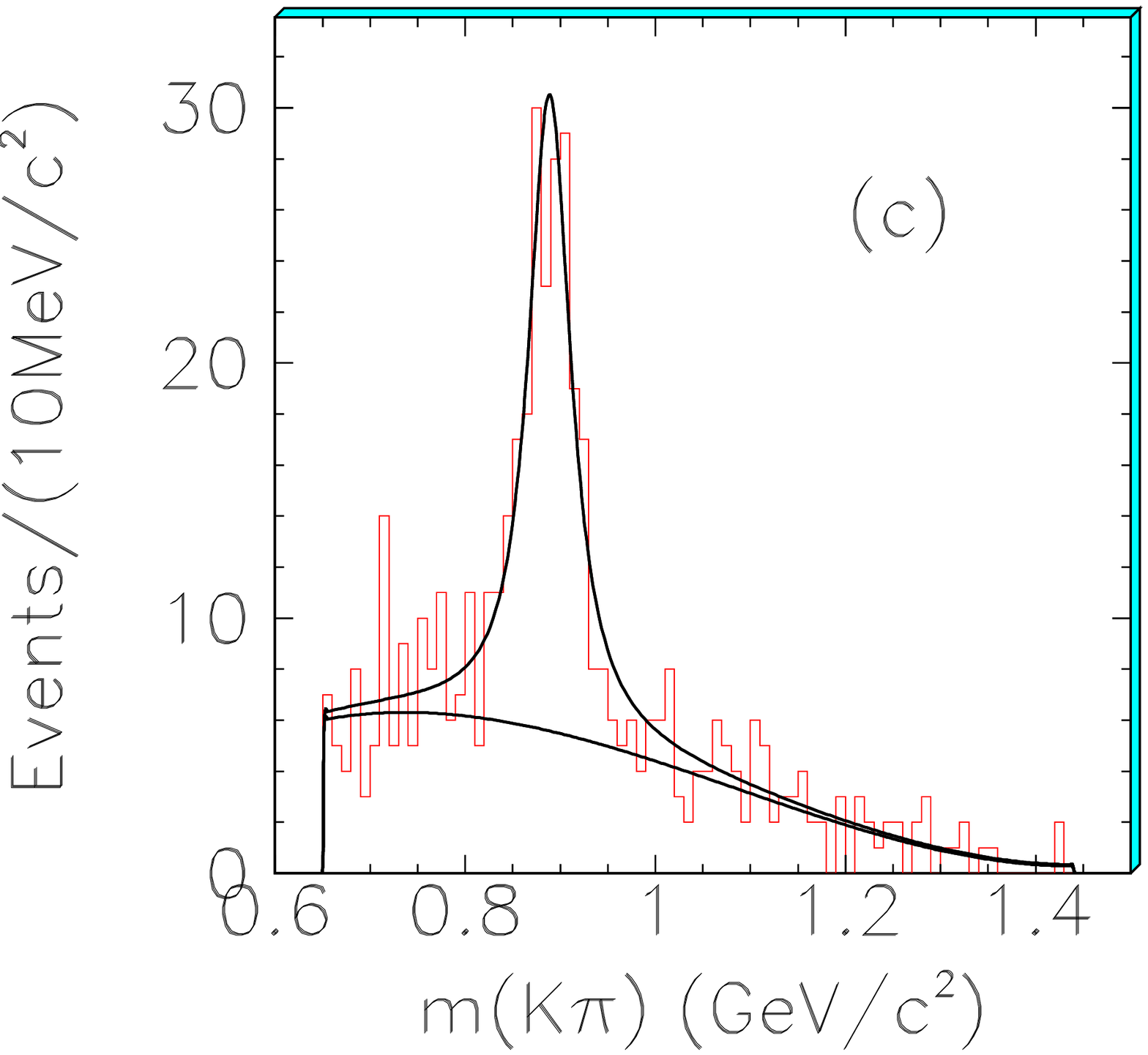}
\caption{(a) The scatter plot of $m_{K^{0}_{S}\pi}$ versus $m_{K\pi}$,
(b) the scatter plot of $m_{\pi^{+}\pi^{-}}$ versus $m_{K^{+}K^{-}}$,
and (c) the combined $K^{0}_{S} \pi^{\mp}$ and $K^{\pm} \pi^{\mp}$
invariant mass distributions for events in the signal region (box 1)
for $J/\psi \rightarrow \phi K^{0}_{S} K^{\pm} \pi^{\mp}$ candidate
events. The curves are the results of the fit described in the text.}
\label{fig:k892-phik892k-kksp}
\end{figure}

To remove most non-$\phi$ background events, the $K^{+}K^{-}$
combination closest to the $\phi$ mass is required to satisfy
$|m_{K^{+}K^{-}}-m_{\phi}|<0.015$ GeV/c$^{2}$. The scatter plot of
$m_{K^{0}_{S}\pi}$ versus $m_{K\pi}$ for candidate events is shown in
Fig. \ref{fig:k892-phik892k-kksp} (a), where the events in the cross
band correspond to the $K^{*}$ signal.

The scatter plot of $m_{\pi^{+}\pi^{-}}$ versus $m_{K^{+}K^{-}}$ is
shown in Figure~\ref{fig:k892-phik892k-kksp} (b), and there is an
accumulation of events in the $\phi$ and $K^{0}_{S}$ cross
bands. Figure~\ref{fig:k892-phik892k-kksp} (c) shows the combined
$K^{0}_{S} \pi^{\mp}$ and $K^{\pm} \pi^{\mp}$ mass
spectrum for events in the
signal region (box 1 in Fig.~\ref{fig:k892-phik892k-kksp} (b)), which
is defined as $|m_{\pi^{+}\pi^{-}}-m_{K^{0}_{S}}|<0.015$ GeV/$c^{2}$
and $|m_{K^{+}K^{-}}-m_{\phi}|<0.015$ GeV/$c^{2}$. A fit yields $194.8
\pm 25.0$ $K^{*}$ events.

The same method as used for the $J/\psi \rightarrow \omega
K^{*}\bar{K} +c.c.\rightarrow \omega K^{0}_{S}K^{\pm}\pi^{\mp}$
analysis is used to estimate the number of background events in the
signal region, and $N_{bg}=(10.0\pm6.6)$, which is neglected in the
branching fraction determination.

 The detection efficiency of $J/\psi \rightarrow \phi
K^{*}\bar{K}+c.c.$ in this decay is $1.42\%$, and its branching
fraction is determined to be
\begin{eqnarray}
B(J/\psi \rightarrow \phi K^{*}\bar{K}+c.c.)
& = &(2.08 \pm 0.27) \times 10^{-3},\nonumber
\end{eqnarray}
where the error is statistical only.

\subsubsection{\bf $J/\psi \to \phi X(1440) \rightarrow \phi K^{0}_{S} K^{\pm} \pi^{\mp}$}

\begin{figure}[htbp]
  \centering
\includegraphics[width=0.35\textwidth]{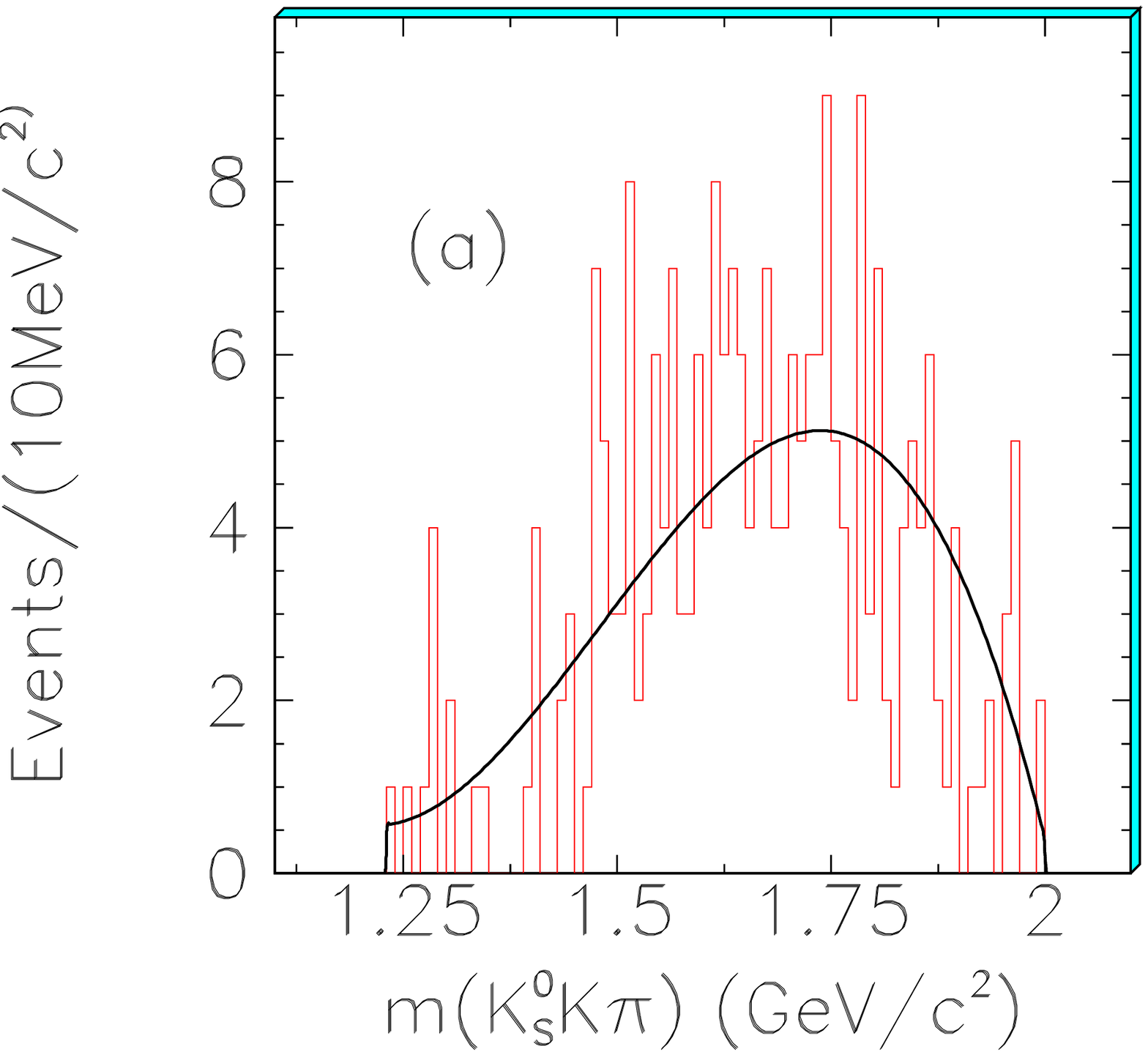}
\includegraphics[width=0.35\textwidth]{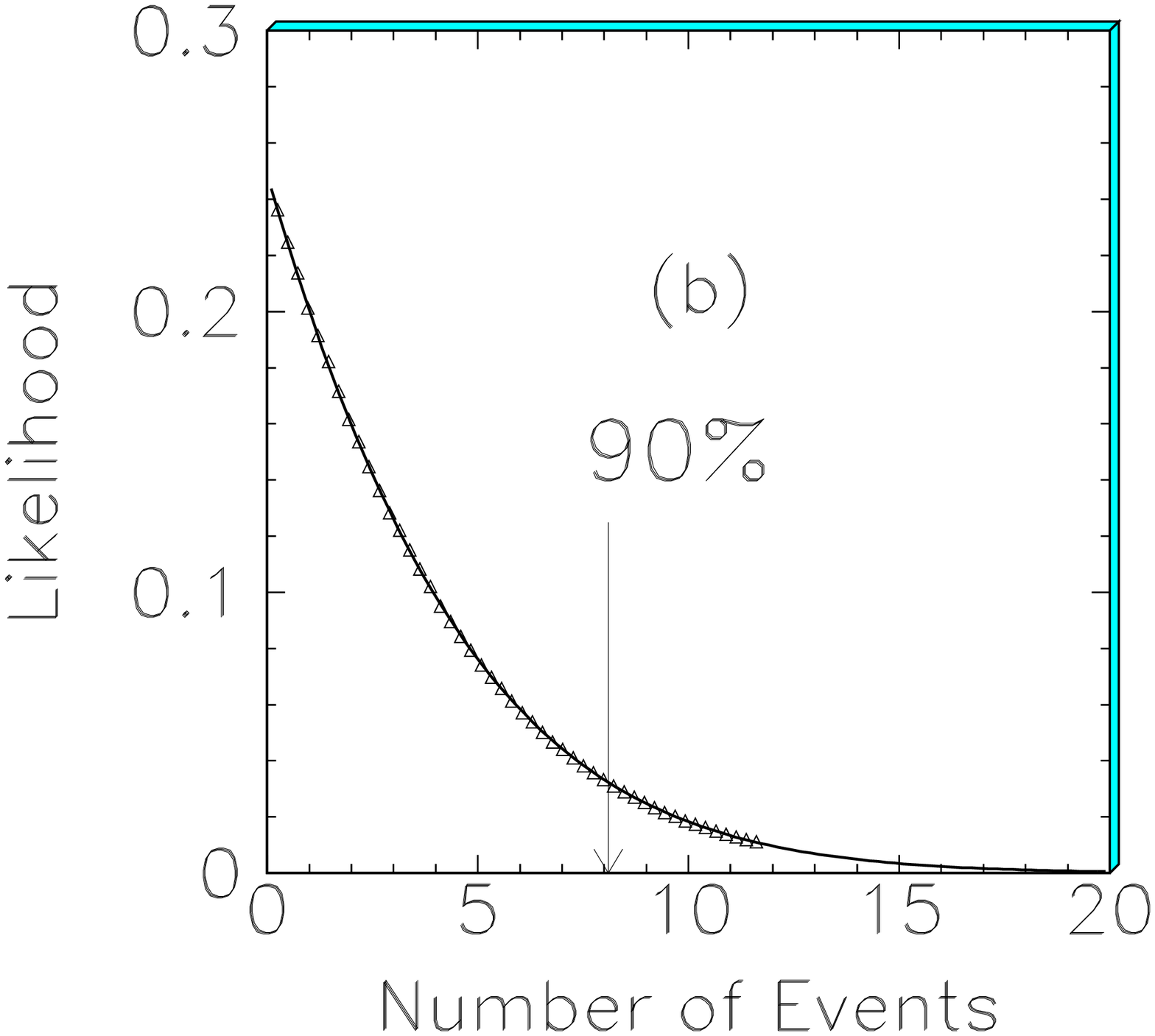}
  \caption{(a) The $K^{0}_{S} K^{\pm} \pi^{\mp}$ invariant mass
  recoiling against the $\phi$, and (b) the number of events of X(1440).  The curve in (a) is a
  third order polynomial to describe the background, and the observed
  number of events at the $90\%$ confidence level using a Bayesian
  method is indicated by the arrow in (b).}
           \label{fig:x1440-phikksp}
\end{figure}

The distribution of $K^{0}_{S} K^{\pm} \pi^{\mp}$ invariant mass
recoiling against the $\phi$ signal is shown in
Fig. \ref{fig:x1440-phikksp} (a), and there is no evidence for
$X(1440)$.  The upper limit on the number of the observed events at
the $90\%$ C.L. is $8.1$~\cite{pdg2006}. The likelihood
distribution and the $90\%$ C.L.  limit are shown in
Fig. \ref{fig:x1440-phikksp} (b). The likelihood values for the number
of events are obtained by fitting the $K^{0}_{S} K^{\pm} \pi^{\mp}$
distributions with a X(1440) signal plus a third-order polynomial
background. Its mass and width are fixed to those in the decay $J/\psi
\rightarrow \omega X(1440)\rightarrow \omega K^{0}_{S}
K^{\pm}\pi^{\mp}$.

 The detection efficiency is $2.53\%$, and
 the upper limit on the branching fraction
at the $90\%$ C.L. is:
\begin{eqnarray}
B(J/\psi\rightarrow \phi X(1440) \rightarrow
\phi K^{0}_{S}K^{+}\pi^{-}+c.c.)<1.93 \times 10^{-5}.
\end{eqnarray}

\subsection{\bf $J/\psi \rightarrow \phi K^{+} K^{-} \pi^{0}$}

\begin{figure}[htbp]
  \centering
\includegraphics[width=0.35\textwidth]{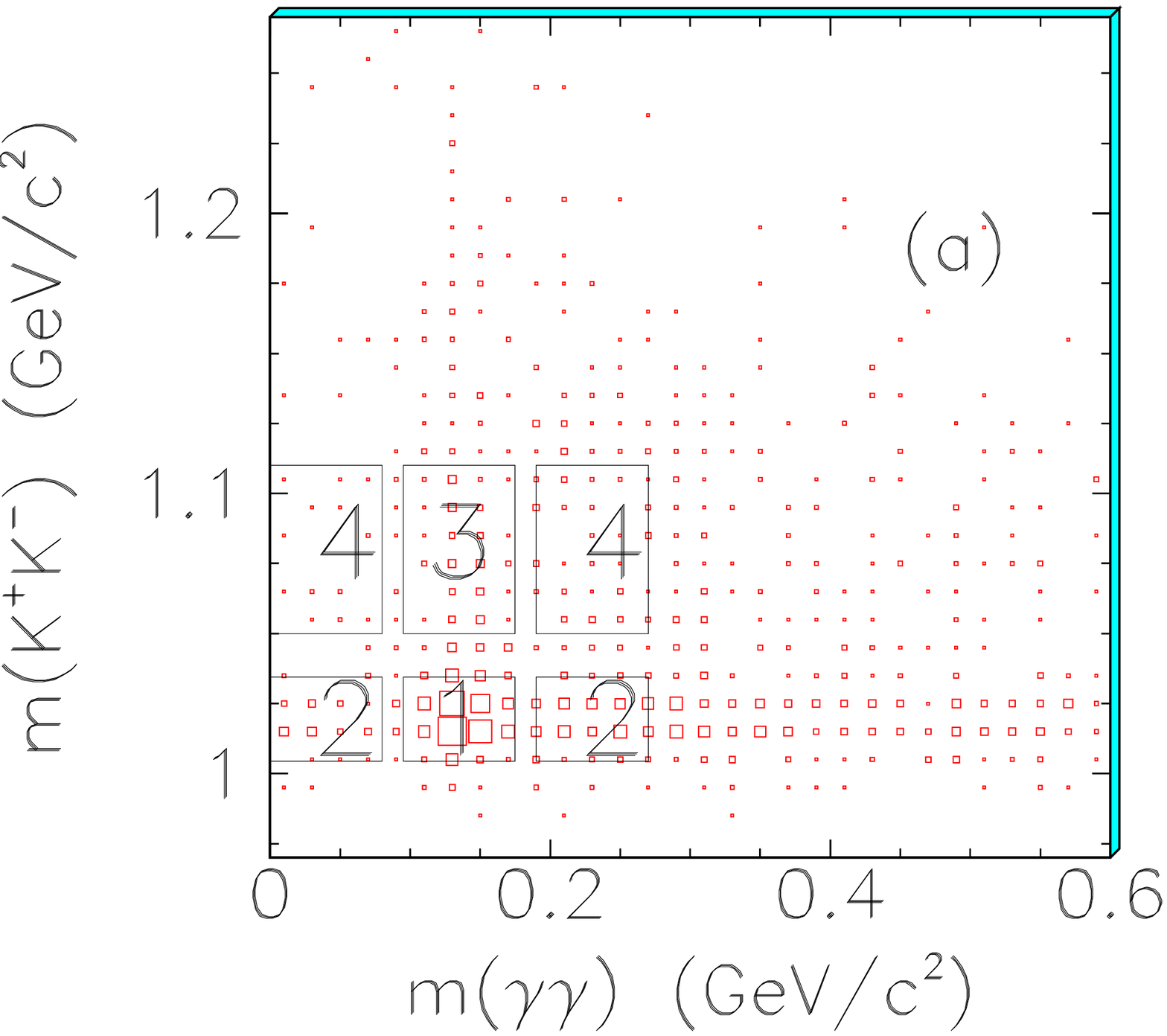}
\includegraphics[width=0.35\textwidth]{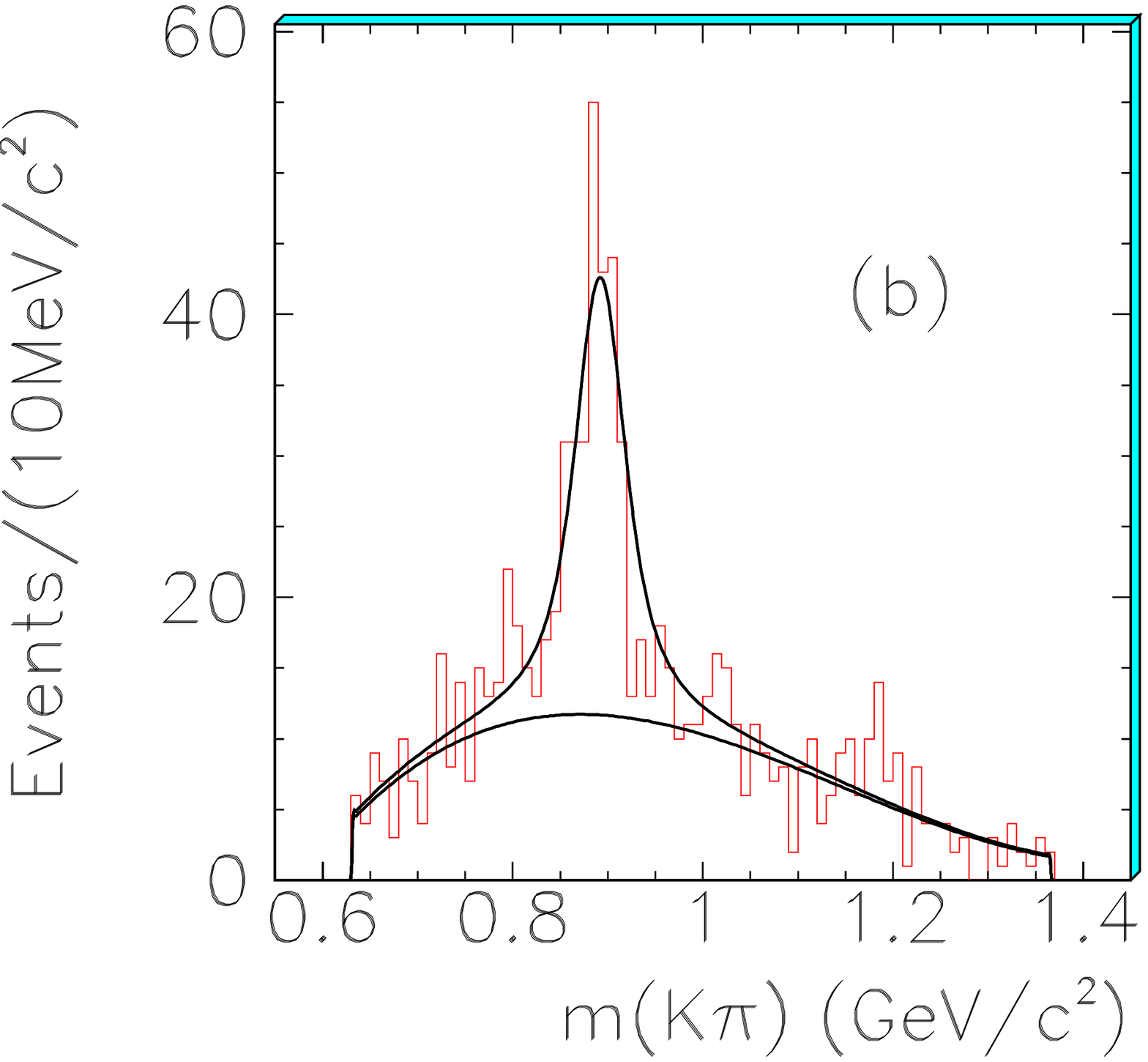}
  \caption{(a) The scatter plot of $m_{\gamma\gamma}$ versus
  $m_{K^{+}K^{-}}$, and (b) the $K^{\pm}\pi^{0}$ invariant mass
  distribution for events in the
  signal region (box 1) for $J/\psi \rightarrow \gamma \gamma
  2(K^{+}K^{-})$ candidate events with two entries per event. The
  curves are the results of the fit described in the text.}
           \label{fig:scatter-phik892k-phi-pi0}
\end{figure}

At least three charged tracks must be identified as kaons. A 4C-fit is
applied under the hypothesis $J/\psi \rightarrow \gamma \gamma
2(K^{+}K^{-})$, and $\chi^{2}<16$ is required. To reject possible
background events from $J/\psi \rightarrow \gamma 2(K^{+}K^{-})$,
the $\chi^{2}$ of the 4C fit for $J/\psi \rightarrow \gamma\gamma
2(K^{+}K^{-})$ is required to be less than that for the $\gamma
2(K^{+}K^{-})$ hypothesis. There are four possible ways to combine
the oppositely charged kaons in forming the $\phi$, and the
$K^{+}K^{-}$ combination closest to the $\phi$ mass is chosen.
Figure~\ref{fig:scatter-phik892k-phi-pi0} (a) shows the scatter plot of
$m_{\gamma \gamma}$ versus $m_{K^{+}K^{-}}$, and clear $\phi$ and
$\pi^{0}$ signals are seen.

\subsubsection{\bf $J/\psi \to \phi K^{*\pm} K^{\mp} \rightarrow \phi K^{+} K^{-} \pi^{0}$}

Figure \ref{fig:scatter-phik892k-phi-pi0} (b) shows the
$K^{\pm}\pi^{0}$ combined mass spectrum for events in the signal
region (box 1 in Fig.~\ref{fig:scatter-phik892k-phi-pi0} (a) ), which
is defined as $|m_{K^{+}K^{-}}-m_{\phi}|<0.015$ GeV/$c^{2}$ and
$|m_{\gamma \gamma}-m_{\pi^{0}}|<0.04$ GeV/$c^{2}$, and a clear
$K^{*\pm}$ signal is seen. It is fitted with a BW, whose mass and
width are fixed to those of $K^{*\pm}$ in the PDG, plus a third-order
polynomial. The number of $K^{*\pm}$ events from the fit is $277.8 \pm
27.7$.  The sidebands are used as before to estimate the number of the
corresponding background events in the signal region, and the result
is $N_{bg}=(40.1\pm
10.1)$.

 After subtracting the above background and incorporating the
efficiency of $1.71\%$ from MC simulation, the branching fraction is
determined to be
\begin{eqnarray}
B(J/\psi \rightarrow \phi K^{*}\bar{K}+c.c. )
& = &(2.96 \pm 0.37) \times 10^{-3},\nonumber
\end{eqnarray}
where the error is statistical only.

\subsubsection{\bf $J/\psi \to \phi X(1440) \rightarrow \phi K^{+} K^{-} \pi^{0}$}

\begin{figure}[htbp]
  \centering
\includegraphics[width=0.35\textwidth]{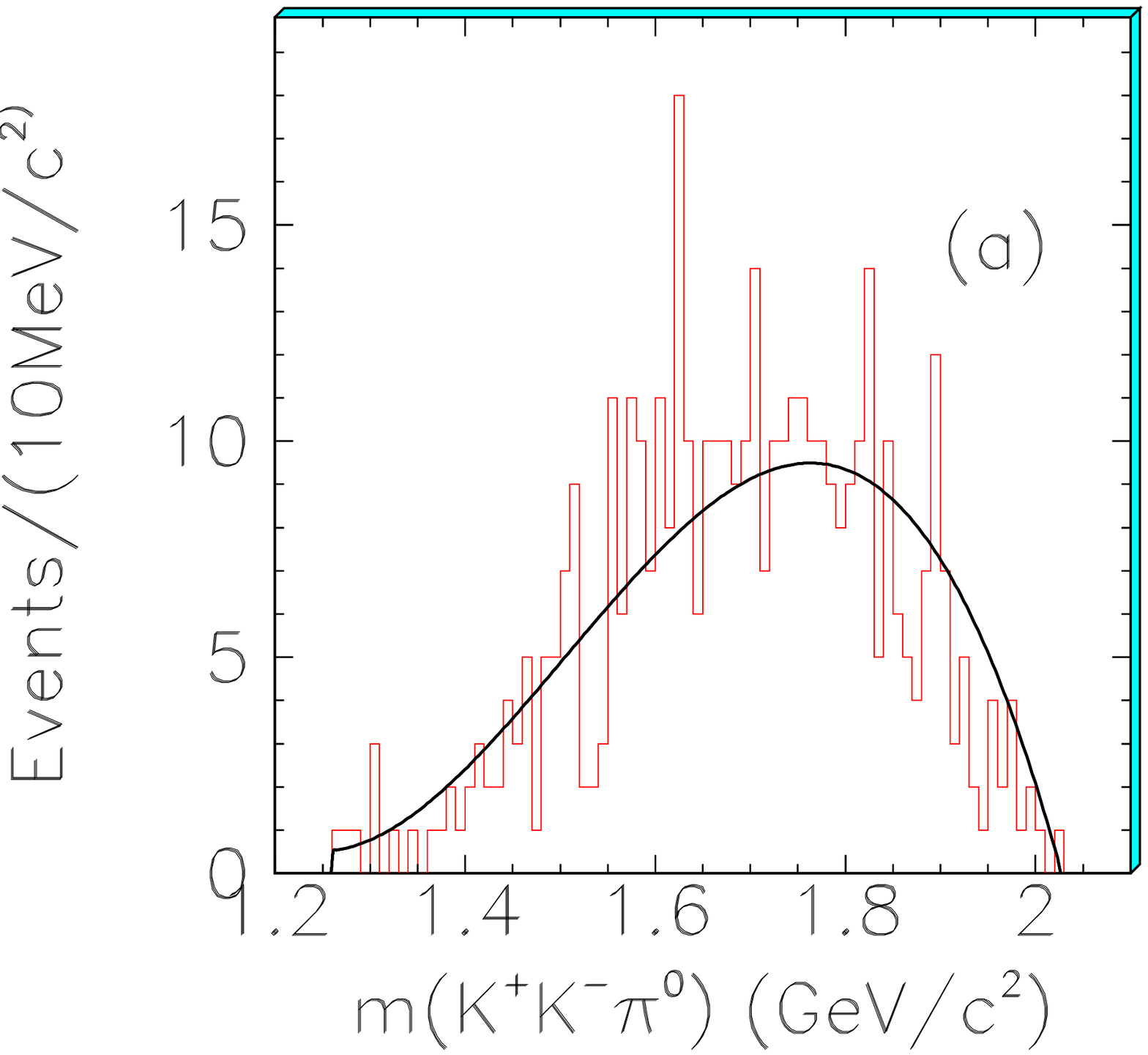}
\includegraphics[width=0.35\textwidth]{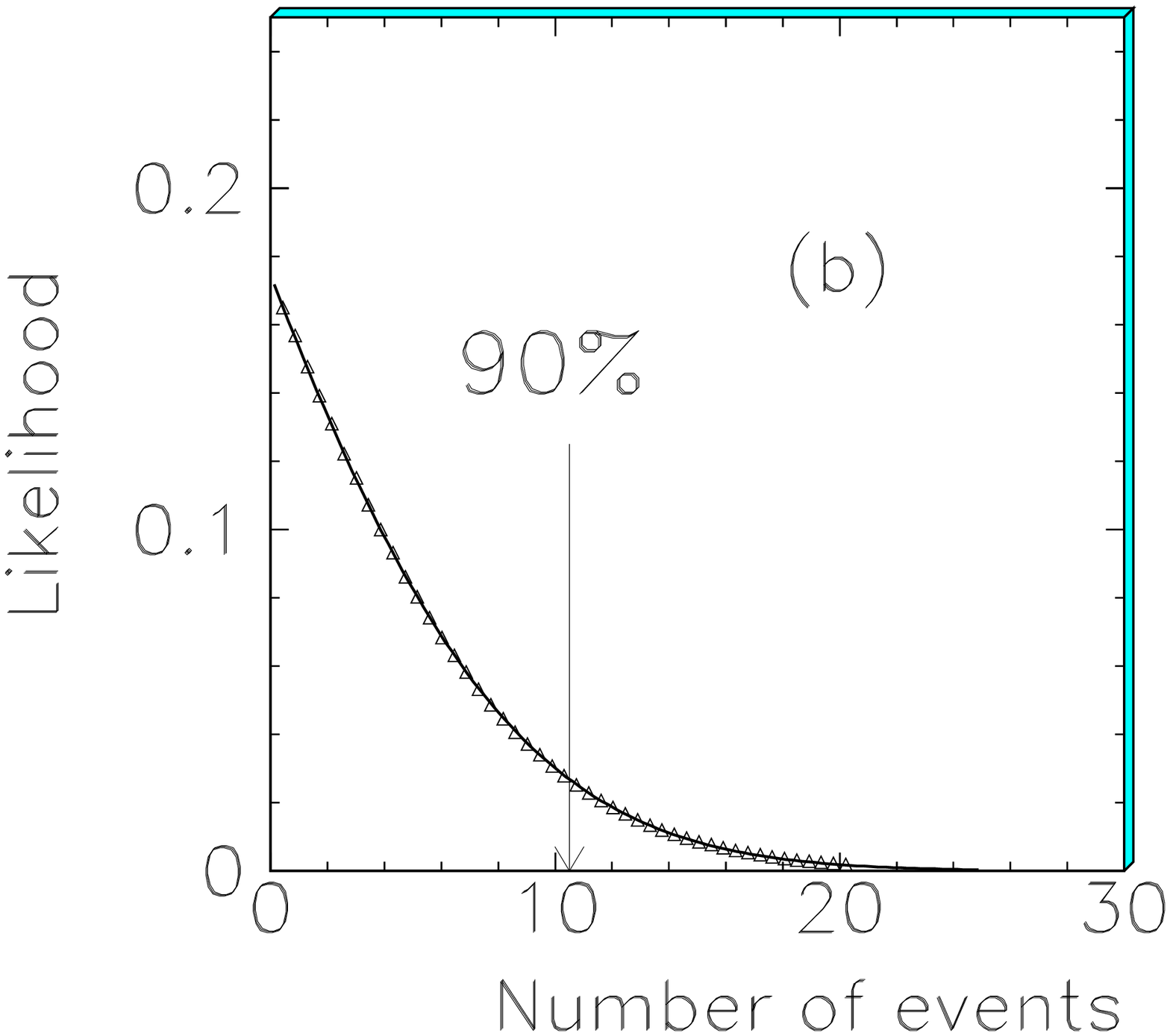}
\caption{(a) The $K^{+} K^{-} \pi^{0}$ invariant mass
recoiling against the $\phi$, and (b) the number of events of X(1440). The curve in (a) is the third-order
polynomial to describe the background, and the observed number of
events at the $90\%$ confidence level using a Bayesian method is indicated
by arrow in (b).} \label{fig:x1440-phikkp0}
\end{figure}

The distribution of $K^{+} K^{-} \pi^{0}$ invariant mass recoiling
against the $\phi$ is shown in Fig. \ref{fig:x1440-phikkp0} (a). No
evidence for the $X(1440)$ is observed near $1440$ MeV/$c^{2}$. The
upper limit on the number of the observed events at the $90\%$ C.L. is
$10.5$~\cite{pdg2006}. The likelihood distribution and the
$90\%$ C.L.  limit are shown in Fig. \ref{fig:x1440-phikkp0} (b). The
likelihood values for the number of events are obtained by fitting the
$K^{+} K^{-} \pi^{0}$ distributions with a X(1440) signal, whose mass
and width are fixed to those of the decay $J/\psi \rightarrow \omega
K^{+} K^{-}\pi^{0}$, plus a third-order background polynomial.
The detection efficiency is $2.49\%$, and the upper limit on the
branching fraction at the $90\%$ C.L. is:
\begin{eqnarray}
B( J/\psi \rightarrow \phi X(1440) \rightarrow \phi
K^{+}K^{-}\pi^{0}) < 1.71 \times 10^{-5}.
\end{eqnarray}

\section{Systematic errors }

In this analysis, the systematic errors on the branching fractions
mainly come from following sources:
\begin{itemize}

\item MDC tracking efficiency\\

 The MDC tracking efficiency is measured from clean
channels like $J/\psi \rightarrow \Lambda \bar{\Lambda}$ and
$\psi(2S) \rightarrow \pi^{+} \pi^{-} J/\psi$ with $J/\psi
\rightarrow \mu^{+} \mu^{-}$. It is found that the MC
simulation agrees with data within $1\%-2\%$ for each charged track.
Therefore, $12\%$ is taken as the
systematic error on the tracking efficiency for the channels with six charged tracks and
$8\%$ for the channels with four charged tracks in the final states.\\

\item Photon detection efficiency \\

The photon detection efficiency is studied from $J/\psi \rightarrow
\rho^{0} \pi^{0}$ events~\cite{lishm2004}. The results indicate that
the difference between data and MC simulation is less than $2\%$ for
each photon. Therefore, $4\%$ is taken to be the systematic error from the
photon efficiency for the channels with two photons and $8\%$ for the
channels with four photons in the final states.\\

\item PID \\

The PID efficiency of the kaon is studied with $J/\psi \rightarrow K^{+}
K^{-} \pi^{0}$ events. The average PID efficiency difference between data and MC is found to be
less than $2\%$.  In this paper, $2\%$, $4\%$, and $6\%$ are
conservatively taken as the systematic errors on PID efficiency for
the channels with one, two, and three identified kaons,
respectively. \\

\item $K^{0}_{S}$ reconstruction \\

The $K^{0}_{S}$ secondary vertex reconstruction is checked using
$J/\psi \rightarrow K^{*\pm}K^{\mp} (K^{*\pm} \rightarrow K^{0}_{S}
\pi^{\pm})$ events.  It is found that the difference of the efficiency
between data and MC simulation is $2.8\%$, which is taken to be the systematic
error from the $K^{0}_{S}$ secondary vertex reconstruction. \\

\item Intermediate decay branching fractions\\

The branching fractions for $\eta \rightarrow \pi^{+}\pi^{-}\pi^{0}$,
$\omega \rightarrow \pi^{+}\pi^{-}\pi^{0}$, and $\phi \rightarrow
K^{+} K^{-}$ are taken from the PDG \cite{pdg2006}, and the errors on
these branching fractions are included as systematic errors in our measurements.
The error on the $K^{0}_{S} \rightarrow \pi^{+}\pi^{-}$ branching
fraction is neglected in this analysis.\\

\item Kinematic fit\\

Kinematic fits are used to reduce backgrounds. Using the same method as
in Ref.~\cite{rhopi2004}, the decay modes $J/\psi \rightarrow
3(\pi^{+}\pi^{-})\pi^{0}$, $J/\psi \rightarrow
2(\pi^{+}\pi^{-})\pi^{0}$, and $J/\psi \rightarrow 3(\pi^{+}\pi^{-})$
are studied~\cite{pi46eta} in order to estimate the corresponding systematic
error. The kinematic fit efficiency differences between data and
MC are $5.5\%$, $4.3\%$, and $8.7\%$, respectively. The efficiency
difference between data and MC for the 6C-kinematic fit to $\eta_{C}
\rightarrow \omega \omega \rightarrow 2(\pi^{+}\pi^{-}\pi^{0})$ is
about $10\%$~\cite{zhounf2005}. Since the decays in this analysis are
similar to the above decays, these systematic errors are taken as the
corresponding systematic errors.\\

\item Background uncertainty\\

The background uncertainties come from the uncertainties associated
with the estimation of the sideband backgrounds, the events from other
background channels, as well as the uncertainties of background
shapes, different fit ranges, and different binning.  Therefore, the
statistical error in the estimated number of background events, the
largest difference from changing background shape, the difference from
changing the fit range, the difference of changing fit binning, and
some ignored backgrounds events are taken as the systematic errors
due to the background uncertainty.\\

\item MC generator \\

There may be interference between charged and neutral $K^{*}$ modes,
which is not included in the MC generator.  In the sample of $J/\psi
\to \omega K^{*}\bar{K}+c.c. \rightarrow \omega
K^{0}_{S}K^{\pm}\pi^{\mp}$ decays, the $K^{\pm}\pi^{\mp}$ mass
distribution in the
$K^{*\pm}$ sideband and signal regions and the $K^{0}_{S}\pi$
mass distributions in the $K^{*0}$ sideband and signal regions from the scatter
plot of $m_{K^{\pm}\pi^{\mp}}$ versus $m_{K^{0}_{S}\pi^{\mp}}$, are studied in
real data and MC simulation. It is found that the difference between data and
MC sample is $5.1\%$, so $5.1\%$ is taken as the systematic error from
the MC
model.\\

\item  Number of $J/\psi$ events \\

The number of $J/\psi$ is $(57.7\pm2.7) \times 10^{6}$, determined
from $J/\psi$ inclusive four-prong events~\cite{fangss2003}.  The
uncertainty is taken as a systematic error in the branching ratio
measurement.
\end{itemize}
Table \ref{table:systematic-error} and Table \ref{table:x1440-systematic-error} list the systematic errors from all above sources,
and the total systematic error is the sum of them added in quadrature.

\begin{table}
\centering
  \caption{Systematic errors in $B(J/\psi \rightarrow \{\eta,\omega,\phi\} K \bar{K} \pi)$.}
  \label{table:systematic-error}
 \begin{tabular}{l|c|c|c|c|c|c|c}\hline\hline
$J/\psi \rightarrow$ & $\eta K^{0}_{S}K^{\pm}\pi^{\mp}$ & $\omega K^{0}_{S}K^{\pm}\pi^{\mp}$
& $\omega K^{*}K\rightarrow$ &  $\omega K^{*}K\rightarrow$
&  $\phi K^{0}_{S}K^{\pm}\pi^{\mp}$ & $\phi K^{*}K\rightarrow$ & $\phi K^{*}K\rightarrow$ \\
    & &  & $\omega K^{0}_{S}K^{\pm}\pi^{\mp}$ & $\omega K^{+}K^{-}\pi^{0}$
    &  & $\phi K^{0}_{S}K^{\pm}\pi^{\mp}$ & $\phi K^{+}K^{-}\pi^{0}$ \\\hline
Error source &  \multicolumn{7}{c}{relative error $(\%)$} \\\hline
MDC tracking         &  12 & 12 & 12      & 8  &
12 & 12  & 8 \\\hline photon efficiency &
4 & 4 & 4   & 8  & - & -   & 4
\\\hline Particle ID       &   2 & 2 & 2   & 2  &
4 & 4   & 6\\\hline $K^{0}_{S}$ 2nd vertex &
2.8 & 2.8 & 2.8 & - & 2.8& 2.8  & -\\\hline
intermediate decays    & 1.8 & 0.8 & 0.8 & 0.8 &
1.2 & 1.2 & 1.2 \\\hline kinematic fit
& 5.5 & 5.5 & 5.5 & 10  & 8.2 & 8.2  &
4.3\\\hline Back. uncertainty      & 1.8 & 2.0 &  5.9   & 11.7 & 8.4
& 6.1 & 7.3 \\\hline
 MC statistic           & 1.9 & 1.3 &   2.3  & 3.2 & 3.2 & 3.4 & 2.2 \\\hline
 MC model & - &  -   & 5.1 &  5.1 & - & 5.1 & 5.1 \\\hline
Number of $J/\psi$ events &  \multicolumn{7}{c}{4.7} \\\hline total
Sum & 15.4 & 15.2 & 17.1 & 20.7  & 18.4 &18.3 &
15.6\\\hline\hline
\end{tabular}
\end{table}

\begin{table}
  \centering
    \caption{Systematic errors in $B(J/\psi \rightarrow \{\omega,\phi\} X(1440)
  \rightarrow \{\omega,\phi\}K \bar{K} \pi)$.}
  \label{table:x1440-systematic-error}
  \begin{tabular}{l|c|c|c|c}\hline\hline
$J/\psi \rightarrow$ &$\omega X(1440)$ & $\omega X(1440)$ & $\phi X(1440)$ & $\phi X(1440)$ \\
                     & $\rightarrow \omega K^{0}_{S}K^{\pm}\pi^{\mp}$ & $\rightarrow \omega K^{+}K^{-}\pi^{0}$
                     & $\rightarrow \phi K^{0}_{S}K^{\pm}\pi^{\mp}$   & $\rightarrow \phi K^{+}K^{-}\pi^{0}$\\\hline
%final state & $2(\pi^{+}\pi^{-})K^{\pm}\pi^{\mp}2\gamma$ & $\pi^{+}\pi^{-}K^{+}K^{-}4\gamma$
%            & $\pi^{+}\pi^{-}K^{+}K^{-}K^{\pm}\pi^{\mp}$ & $2(K^{+}K^{-})2\gamma$  \\\hline
Error source &  \multicolumn{4}{c}{relative error $(\%)$} \\\hline
MDC tracking         &  12      & 8  & 12  & 8 \\\hline
photon efficiency &  4       & 8  &  -   & 4 \\\hline
Particle ID       &  2       & 2  & 4   & 6 \\\hline
$K^{0}_{S}$ 2nd vertex & 2.8 & -  & 2.8 & - \\\hline
intermediate decays    & 0.8 & 0.8 & 1.3 & 1.2 \\\hline
kinematic fit          & 5.5 & 10  & 8.2 & 4.3 \\\hline
%Back. uncertainty      & 5.9 & 2.9   & -&  -  \\\hline   ! old fit
MC statistic           & 2.7 & 2.6   & 0.8   & 0.8 \\\hline
Back. uncertainty      & 6.4 & 10.9   & -&  -  \\\hline
Number of $J/\psi$ events  &  \multicolumn{4}{c}{4.7} \\\hline
Sum & 16.6 & 19.6  & 16.2     & 12.6 \\\hline\hline
\end{tabular}
\end{table}

\section{\bf Results}

Table \ref{table:wphikksp-bes} lists the branching fractions of
$J/\psi\rightarrow\{\eta,\omega,\phi\} K^{0}_{S}K^{\pm}\pi^{\mp}$,
$J/\psi\rightarrow \{\omega,\phi\} K^{*}\bar{K}+c.c.$ from different
decay modes.  These branching fractions are somewhat larger than those of other
experiments in Table \ref{table:wphikksp-pdg}~\cite{wkkpmark3}~\cite{dm288} but they are still consistent within errors.
The branching fraction for $J/\psi\rightarrow \eta
K^{0}_{S}K^{\pm}\pi^{\mp}$ is measured for the first time. In the
invariant mass spectra of $K^{0}_{S}K^{\pm}\pi^{\mp}$ and
$K^{+}K^{-}\pi^{0}$ recoiling against the $\omega$, the resonance
at $1.44$ GeV/$c^{2}$ is observed, with the mass, width, and branching fractions
listed in Table \ref{table:x1440-branching-ratio}; while in the invariant mass
spectra of $K^{0}_{S}K^{\pm}\pi^{\mp}$ and $K^{+}K^{-}\pi^{0}$
recoiling against the $\phi$, no significant structure near 1.44
GeV/c$^2$ is seen and an upper limits on the $J/\psi$ decay branching
fractions at
the $90\%$ C.L. are given in Table \ref{table:x1440-branching-ratio}.

\begin{table}
  \centering
  \caption{The branching fractions of $J/\psi$ decays in BESII.}
  \label{table:wphikksp-bes}
  \begin{tabular}{l|l|l|c|l}\hline\hline
Decay   &   final state & No. of events  &  efficiency & Branching fraction  ($10^{-4}$)\\\hline

$\omega K^{0}_{S}K^{+}\pi^{-}+c.c.$  &
$(\pi^{+}\pi^{-}\pi^{0})K^{0}_{S}K^{\pm}\pi^{\mp}$&  $1971.7\pm 41.4$ & $1.48\%$ &
$37.7\pm0.8\pm5.8$ \\\hline $\eta K^{0}_{S}K^{+}\pi^{-}+c.c.$ &
$(\pi^{+}\pi^{-}\pi^{0})K^{0}_{S}K^{\pm}\pi^{\mp}$ & $231.6\pm 23.1$   & $1.18\%$  &
$21.8\pm 2.2 \pm 3.4$ \\\hline $\omega K^{*}\bar{K}+c.c.$  & $(\pi^{+}\pi^{-}\pi^{0})K^{0}_{S}K^{\pm}\pi^{\mp}$
& $898.7\pm97.7$ & $1.23\%$
& $62.0 \pm 6.8 \pm10.6$ \\
                  &  $(\pi^{+}\pi^{-}\pi^{0})K^{+}K^{-}\pi^{0}$  &  $175.6\pm27.4$  & $0.32\%$ & $65.3\pm10.2\pm13.5$ \\\hline
$\phi K^{0}_{S}K^{+}\pi^{-}+c.c.$  &  $(K^{+}K^{-})K^{0}_{S}K^{\pm}\pi^{\mp}$ &  $227.1\pm19.0$ & $1.56\%$ & $7.4\pm0.6 \pm1.4$\\\hline
$\phi K^{*}\bar{K}+c.c.$   &  $(K^{+}K^{-})K^{0}_{S}K^{\pm}\pi^{\mp}$ &  $194.8\pm25.0$  & $1.42\%$ &  $20.8\pm2.7\pm3.9 $ \\
                 &  $(K^{+}K^{-})K^{+}K^{-}\pi^{0}$  &  $237.7\pm29.5$  & $1.71\% $ &  $29.6\pm3.7\pm4.7$ \\\hline\hline
\end{tabular}
\end{table}
\begin{table}
  \centering
    \caption{The branching fractions of $J/\psi$ decays from MarkIII~\cite{wkkpmark3} and DM2~\cite{dm288} Collaborations}.
  \label{table:wphikksp-pdg}
  \begin{tabular}{l|l|l|c}\hline\hline
    & Decay   & final state &   Branching fraction  ($10^{-4}$)\\\hline
MarkIII  & $\omega K^{0}_{S}K^{+}\pi^{-}+c.c.$ & $(\pi^{+}\pi^{-}\pi^{0})K^{0}_{S}K^{\pm}\pi^{\mp}$  & $29.5\pm1.4\pm7.0$ \\\cline{2-4}
       & $\omega K^{*}\bar{K}+c.c.$ &  $(\pi^{+}\pi^{-}\pi^{0})K^{0}_{S}K^{\pm}\pi^{\mp}$ &  $53\pm14\pm14$ \\
       &    &  $(\pi^{+}\pi^{-}\pi^{0})K^{+}K^{-}\pi^{0}$  & \\\cline{2-4}
       & $\phi K^{0}_{S}K^{+}\pi^{-}+c.c.$ & $(K^{0}_{S}K^{0}_{L}))K^{0}_{S}K^{\pm}\pi^{\mp}$ & $7.0\pm0.6\pm1.0$ \\
       &      & $(K^{+}K^{-})K^{0}_{S}K^{+}\pi^{-}+c.c.$ &  \\\hline
DM2   &$\phi K^{0}_{S}K^{+}\pi^{-}+c.c.$ &  $(K^{+}K^{-})K^{0}_{S}K^{\pm}\pi^{\mp}$  & $7.4\pm0.9\pm1.1$  \\\cline{2-4}
       & $\phi K^{*}\bar{K}+c.c.$ &  $(K^{+}K^{-})K^{0}_{S}K^{\pm}\pi^{\mp}$  & $20.8\pm2.7\pm3.7$ \\\hline\hline
\end{tabular}
\end{table}

%A mass enhancement in the $K\bar{K}\pi$ system recoiling against the
%$\omega$ is observed with the mass, width, and branching fractions
%listed in Table \ref{table:x1440-branching-ratio}.  No evidence of a
%mass enhancement in the $K \bar{K} \pi$ invariant mass recoiling
%against the $\phi$ is seen, and upper limits on branching fractions at
%the $90\%$ C.L. are given in Table \ref{table:x1440-branching-ratio}.

\begin{table}
  \centering
  \caption{The mass, width, and branching fractions of $J/\psi$ decays into ${\{\omega,\phi\}} X(1440)$.}
  \label{table:x1440-branching-ratio}
  \begin{tabular}{l|l}\hline\hline
$J/\psi \rightarrow \omega X(1440)$    & $J/\psi \rightarrow \omega X(1440)$  \\
($X\rightarrow K^{0}_{S}K^{+}\pi^{-}+c.c.$) & ($X\rightarrow K^{+}K^{-}\pi^{0}$)  \\\hline
$M=1437.6\pm 3.2$ MeV/$c^{2}$    & $M=1445.9\pm 5.7 $ MeV/$c^{2}$       \\
$\Gamma=48.9 \pm 9.0$ MeV/$c^{2}$  & $\Gamma=34.2\pm18.5$ MeV/$c^{2}$ \\\hline
\multicolumn{2}{l}{$B( J/\psi\rightarrow \omega X(1440)\rightarrow \omega
K^{0}_{S}K^{+}\pi^{-}+c.c.)=(4.86\pm0.69\pm0.81) \times 10^{-4}$}\\\hline
\multicolumn{2}{l}{$B( J/\psi\rightarrow \omega X(1440) \rightarrow \omega K^{+}K^{-}\pi^{0})~~~~~~~~=
(1.92\pm0.57\pm0.38) \times 10^{-4}$} \\\hline
\multicolumn{2}{l}{$B(J/\psi\rightarrow \phi X(1440) \rightarrow
\phi K^{0}_{S}K^{+}\pi^{-}+c.c.)<1.93 \times 10^{-5}$} ($90\%$ C.L.)\\\hline
\multicolumn{2}{l}{$B( J/\psi \rightarrow \phi X(1440) \rightarrow \phi K^{+}K^{-}\pi^{0})~~~~~~~~< 1.71 \times 10^{-5}$} ($90\%$ C.L.)\\\hline \hline
\end{tabular}
\end{table}

\acknowledgments

The BES collaboration thanks the staff of BEPC and computing
center for their hard efforts. This work is supported in part by
the National Natural Science Foundation of China under contracts
Nos. 10491300, 10225524, 10225525, 10425523, 10625524, 10521003,
the Chinese Academy of Sciences under contract No. KJ 95T-03, the
100 Talents Program of CAS under Contract Nos. U-11, U-24, U-25,
and the Knowledge Innovation Project of CAS under Contract Nos.
U-602, U-34 (IHEP), the National Natural Science Foundation of
China under Contract No. 10225522 (Tsinghua University), and the
Department of Energy under Contract No. DE-FG02-04ER41291 (U.
Hawaii).

\end{document}